\documentclass{aa}  
\usepackage{graphicx}
\usepackage{natbib}
\usepackage{xcolor}
\usepackage{txfonts}
\graphicspath{{./}{figures/}}
\begin{document}

\title{Dynamical study of Geminid formation assuming a rotational instability scenario}

\author{Hangbin Jo
          \inst{1,2}
          \and
          Masateru Ishiguro \inst{1,2}
          }

   \institute{Department of Physics and Astronomy, Seoul National University, 1 Gwanak-ro, Gwanak-gu, Seoul 08826, Republic of Korea
   \and SNU Astronomy Research Center, Department of Physics and Astronomy, Seoul National University, 1 Gwanak-ro, Gwanak-gu, Seoul 08826, Republic of Korea\\
              \email{hangbin9@snu.ac.kr ishiguro@snu.ac.kr}
             }

\date{Received <date> /
Accepted <date>}

\abstract
   {Various ideas have been proposed to explain the formation of the Geminid meteoroid stream from the asteroid (3200) Phaethon. However, little has been studied about the Geminid formation based on the assumption that mass ejection happened from this asteroid via rotational instability.}
   {Here, we present the first dynamical study of the Geminid formation, taking account of low-velocity mass ejection as a result of Phaethon's rotational instability.}
   {We conducted numerical simulations for 1 mm and 1 cm particles ejected in a wide range of ejection epochs (10$^3$--10$^5$ years ago). We computed the minimum orbital intersecting distance (MOID) of the dust particles as the realistic condition, that is, the Earth's radius and the Earth-Moon distance to be observed as the Geminid meteoroid stream.}
   {We found that the low-velocity ejection model produced the Geminid-like meteoroid stream when the dust particles were ejected more than $\sim$2,000 years ago. In this case, close encounters with terrestrial planets would transport some dust particles from the Phaethon orbit (the current MOID is as large as $\sim$460 Earth radius) to the Earth-intersecting orbits. The optimal ejection epoch and the estimated mass were 18\,000 years ago and  $\sim 10^{10} - 10^{14}$ g (<0.1 \% of the Phaethon mass).}
   {Our results suggest that the JAXA's DESTINY\textsuperscript{+} mission has the potential to find evidence of recent rotational instability recorded on the Phaethon's surface.}
   

\maketitle
%

\section{Introduction} \label{sec:intro}

The interplanetary space of the Solar System is populated by interplanetary dust particles (IDPs). Despite the smallest constituents, these particles are important subjects for investigating the evolution of the Solar System's small bodies. Although a significant fraction of these dust particles are linked to cometary origins  \citep{2010ApJ...713..816N}, the question remains open regarding the source and evolution of other dust particles that are unseen due to bias and observational limitations \citep{2020P&SS..19004973L}.

A near-Earth asteroid, (3200) Phaethon, is the target asteroid of JAXA's DESTINY\textsuperscript{+} mission \citep{2018LPI....49.2570A}. It is an interesting specimen for several reasons. Phaethon has a highly eccentric orbit ($e$=0.89) and a small perihelion distance ($q$=0.14 au). Although near-Earth object population models predict that more asteroids should share similar orbital traits, the lack of them raises the possibility that additional mechanisms have degraded objects getting too close to the Sun. It is, therefore, important to investigate the Geminid meteor stream, which is thought to be fragments of Phaethon and could retain clues about the nature of such mechanism. Ever since the connection between Phaethon and the Geminids was made \citep{1983Natur.306..116H,1983IAUC.3881....1W}, there has been extensive research on how Phaethon produced this dust stream. \citet{2010AJ....140.1519J,2013AJ....145..154L,2013ApJ...771L..36J,2017AJ....153...23H}  observed dust activity on Phaethon near its perihelion, but they all commonly found the particles to be of a micron-scale size, and the dust production rate was deemed too low to fuel the Geminid stream. Moreover, recent studies suggest that this present-day activity is predominantly sodium volatilization, while dust production is minimal \citep{2021PSJ.....2..165M, 2023PSJ.....4...70Z}. In comparison, observations of the Geminid shower revealed the presence of mm- and cm-sized dust particles \citep{2010IAUS..263..218B,2017P&SS..143...83B}.

So far, it is still a mystery as to how Phaethon produced a dust stream whose mass could be comparable to Phaethon itself \citep{2017P&SS..143..125R} and how its present-day activity is related to the birth of the Geminids. Several dust ejection mechanisms have been proposed as the cause of the Geminids, with the earliest one being comet-like ice sublimation \citep{1989A&A...225..533G}. However, thermal modeling suggested that there would not be enough ice in Phaethon for such activity to occur \citep{2010AJ....140.1519J,2012AJ....143...66J}. On the contrary, \citet{2019MNRAS.482.4243Y} found that Phaethon could have retained a small amount of ice in the core. Even in this case, the resulting sublimation would be too weak to cause a spontaneous, large-scale ejection event. Asteroidal dust ejection by thermal fracture is also a viable scenario, based on its possible contribution to the particle ejection event on asteroid (101955) Bennu observed by OSIRIS-REx (Origins, Spectral Interpretation, Resource Identification, and Security–Regolith Explorer) \citep{2019Sci...366.3544L,2020JGRE..12506325M} and it has been suggested as the ejection mechanism for Phaethon by \citet{2010AJ....140.1519J} and \citet{2012AJ....143...66J}. However, questions remain as to whether thermal fracture can cause massive mass ejection and, if so, why such a phenomenon has not been observed in modern times. The feasibility of other mechanisms, such as electrostatic lofting \citep{2022Icar..38215022K} and thermal radiation pressure \citep{2021A&A...654A.113B}, have also been proposed for small-sized dust ejection. However, none are suitable explanations for the larger dust sizes populating the Geminids. Lastly, \citet{2020ApJ...892L..22N} hypothesized that YORP  (Yarkovsky–O'Keefe–Radzievskii–Paddack effect)-induced rotational instability may have formed the Geminids. Indeed in recent years, asteroid activity by rotational instability has been discovered in cases such as (6478) Gault \citep{2021ApJ...910L..27L} and 311P/Gibbs \citep{2021AJ....162..268J}. A recent observation indicates that Phaethon's rotation accelerates by $2.1 \times 10^{-6}$ \degr/day \citep{2022DPS....5451407M}. If a similar acceleration happened in the past, rotational instability could be the key to the Geminid formation. Nonetheless, the cause of the Geminid formation and its relation to the current activity on Phaethon is still under debate and not fully understood. To investigate Phaethon and its dust environment in detail, Demonstration and Experiment of Space Technology for INterplanetary voYage, Phaethon fLyby and dUst Science Phaethon fLyby with reUSable probe (DESTINY\textsuperscript{+}) mission is planned for a flyby mission \citep{2018LPI....49.2570A} and will provide us with more observational constraints.

In addition to these observational studies of Phaethon, theoretical approaches have been conducted to trace back the history of the Geminid stream. The initial dust ejection velocities, the most critical assumption for dust evolution, are assumed based on the above-mentioned dust ejection models. In particular, Whipple's formula \citep{1951ApJ...113..464W} is the most widely applied to derive the ejection velocity by assuming a comet-like mechanism. This model yields up to several hundreds of m s$^{-1}$ \citep{1982MNRAS.200..313F,1983MNRAS.205.1155F,1986acm..proc..549H,1986MNRAS.223..479J,1993MNRAS.262..231W,2007MNRAS.375.1371R,2008EM&P..102...95R,2013SoSyR..47..219R,2016MNRAS.456...78R,2017P&SS..143..125R}. Although not an example of Whipple's formula, despite the similar velocity regime, it is worth mentioning that \citet{2015MNRAS.453.1186J} simulated sequential ejections of particles from $10^3 - 10^5$ years with 1/1000 of the orbital velocity at perihelion as the initial velocity ($\sim 200$ m s\textsuperscript{-1}). Thermal fatigue was addressed by \citet{2012MNRAS.423.2254R,2018MNRAS.479.1017R} by using the upper limit ejection velocity of 30 m s$^{-1}$ \citep{2010AJ....140.1519J}, but neither could successfully recreate the Geminid stream. In the case of lower ejection velocities, \citet{2018ApJ...864L...9Y} tested the gravitational escape speed as the initial velocity to investigate the possibility of Phaethon fragments released within three centuries can come into a close encounter with Earth. However, they did not consider the Geminid formation with a realistic criterion, but they addressed whether large fragments would be observable via telescopic observations, applying an enormously large criterion of $\leq$0.02 au from Earth (about 10 times further than the Earth-Moon distance). Meanwhile, \citet{2012MNRAS.423.2254R} used a wide range of ejection velocities extending from escape velocity to 100 m s$^{-1}$. However, the author found no particles reaching less than 0.01 au from Earth ($\sim$230 Earth radius). Similarly, \citet{2023PSJ.....4..109C} explored a velocity range from zero to km s$^{-1}$ scale, assuming that the ejection happened 2000 years ago. Although this study is close to the timescale and velocity regime of our interest, it was done with a relatively small number of particles, leading to the use of a loose distance limit of 0.03 au to determine the particles' interaction with Earth. Despite numerous dynamical studies on the Geminid stream, there is no research on the Geminid formation based on the assumption that Phaethon's rotational instability generated the Geminid meteoroid stream.

Motivated by a more realistic dust ejection study in \citet{2020ApJ...892L..22N} and other associated observations of top-shape asteroids like (162173) Ryugu and (101955) Bennu, we conducted the dynamical simulation that mm- and cm-sized meteoroid particles were ejected from Phaethon with low velocity via rotational instability. We emphasize that our dynamical study is more rigorous and comprehensive than many previous studies in setting the strict condition of the minimum orbital intersecting distance (MOID) and investigating a wide range of ejection epochs (10$^3$ to 10$^5$ years). Based on our numerical results, we discuss the detailed formation mechanism of the Geminids and describe the expectation of the JAXA’s DESTINY\textsuperscript{+} mission.

\section{Method} \label{sec:method}

\begin{table*}
\caption{Orbital elements of Phaethon (JD 2459000.5)}              
\label{inival_table}      
\centering                                      
\begin{tabular}{c c}          
\hline\hline                        
Element      &  Value  \\    
\hline                                   
    Semi-major axis, $a$ [au] & 1.2713678841019822 \\
    Eccentricity, $e$ & 0.8898311206936256     \\
    Inclination, $i$ [\degr] & 22.25951171307781 \\
    Longitude of ascending node, $\Omega$ [\degr]  & 265.2176958865816  \\
    Argument of perihelion, $\omega$ [\degr] & 322.1867098330639 \\
    Mean anomaly, $M$ [\degr] & 228.9572432413808 \\
    Non-gravitational transverse acceleration parameter, $A_2$ [$10^{-15}$ au/day$^2$] & -5.444804809526769 \\
    \hline
\hline                                             
\end{tabular}
\tablefoot{The covariance matrix of the orbital elements can be found in Appendix \ref{app:clones}.}
\end{table*}

We conducted our dynamical simulations using the Mercury6 N-body integrator \citep{1999MNRAS.304..793C} with the following additional equations.

\begin{equation}
\beta = 5.7 \times 10^{-5} Q_\mathrm{PR} / \rho s, \\
\label{eq:1-1}
\end{equation}

\begin{equation}
F_\mathrm{rad} = F_\mathrm{g} \beta \left\{\left[1 - 2 (1 + \xi) \frac{\vec{v} \cdot \vec{r}}{cr}\right]\hat{r} - (1 + \xi) \frac{\vec{v}}{c} \right\},
\label{eq:1-2}
\end{equation}

\noindent and

\begin{equation}
F_\mathrm{PN} = - \frac{3 G^2 {M_{\odot}}^2 m a (1-e^2)}{c^{2} r^{4}} \hat{r}.
\label{eq:2}
\end{equation}

Eq. (\ref{eq:1-1}) defines $\beta$, the ratio of radiation pressure force $F_\mathrm{rad}$ to the Newtonian gravitational force $F_\mathrm{g}$. In Eq. (\ref{eq:1-1}), $Q_\mathrm{PR}$, $\rho$, and $s$ denote the radiation pressure coefficient, particle mass density, and radius, respectively \citep{1979Icar...40....1B}. Eq. (\ref{eq:1-2}) characterizes the radiation pressure and Poynting-Robertson (PR) drag exerted on a particle at heliocentric position $\vec{r}$ and velocity $\vec{v}$, while $c$ is the light speed \citep{1979Icar...40....1B, 1995Icar..116..186L,2012MNRAS.421..943K}. For the solar wind coefficient $\xi$, we assumed 0.3 \citep{1994AREPS..22..553G}. Eq. (\ref{eq:2}) is the post-Newtonian correction formula \citep{thornton2004classical,doi:10.1119/1.1949625}, where $a$, $e$, and $m$ denote the semi-major axis, eccentricity, and mass of the Solar system object. We used Mercury6 for two purposes: (i) to find the position and velocity of Phaethon in the past when dust particles were ejected and (ii) to trace the orbital evolution of ejected dust particles until the present day. We chose the Bulirsch-Stoer algorithm \citep{stoer1980introduction} with an initial time step of 1 day for all our simulations for this study. 

Backward integration of Phaethon is essential to obtain accurate orbital positions and velocities of the dust source at the dust ejection epoch. Mercury6 offers a built-in backward integration feature. The feature works in two steps: the backward integration to the user-defined initial epoch and subsequent forward integration to the final epoch, during which the particle dynamics are recorded. In other words, Mercury6 does a duplicate integration of the same path backward and forward for the sake of output recording. In the case of Phaethon, which is subject to the gravitational perturbation of terrestrial planets, we find that this process amplifies the chaotic nature of Phaethon's orbital history. Therefore, we chose not to use this feature for this process, and instead, the backward integration was realized by forward integrating with reversed velocities. 

The initial positions of the 8 planets and Phaethon are obtained based on the JPL DE431 ephemerides at JD 2459000.5, provided by JPL Horizons\footnote{\url{https://ssd.jpl.nasa.gov/horizons.cgi}}. Using the osculating orbital elements (see Table \ref{inival_table} and \ref{table:covariance}), we created 100 clones from the multivariate normal distribution generated by the covariance matrix of the orbital elements. We integrated the dust trajectories for about $10^5$ yr with an output interval of 1 day and created an “archive” of the orbital evolution of Phaethon and its clones, from which we could extract the orbital position and velocity of Phaethon at any given day within the past $10^5$ yr.

The age of the Geminids has been estimated by various studies with varying methods, such as backward integration of meteoroids \citep{1989A&A...225..533G}, spin-up timescale to observed flickering of fireballs \citep{2002MNRAS.336..559B}, and analysis of meteoroid orbital separation \citep{1999SoSyR..33..224R}. Most of these works indicate that the Geminid age is in the order of $10^3$ years. Therefore, we selected ejection epochs between $\sim 10^3$ and $\sim 10^4$ years ago with $10^3$-year intervals. Specifically, we started with the ejection epoch about 2000 years ago, JD 1740898.5 (April 25, 54), when Phaethon's perihelion distance was smallest. We then selected 8 more epochs in the past with an interval of 365\,000 days. While we recognize that 365\,000 days is not equal to 1000 years, we nonetheless chose this value as an appropriate search interval for approximating the optimal ejection epoch. In actual calculations throughout this study, 1 year is equal to 365.25 days. For a more detailed investigation of the $\sim 10^4$-year-old regime, we also conducted additional simulations for ejection epochs of approximately 15, 18, 20, 25, 30, and 100 thousand years ago. Again, we used the interval of 365\,000 days starting from JD 1740898.5 when selecting the ejection epochs. We only rely on the ejection epoch as our initial setting, resulting in the asteroid's initial position in the orbit effectively being random rather than fixed at perihelion or aphelion. 

The forward integration from the ejection epoch to the present day requires the $\beta$ value and the velocity of the particles, and the initial position of the dust source. We set the $\beta$ value of the dust particles in Eq. (\ref{eq:1-1}), assuming that $Q_\mathrm{PR}$ is unity and mass density is 2.9 g/cm\textsuperscript{3}. We chose particle radii of 1 mm and 1 cm.

\begin{figure}
 \resizebox{\hsize}{!}{\includegraphics{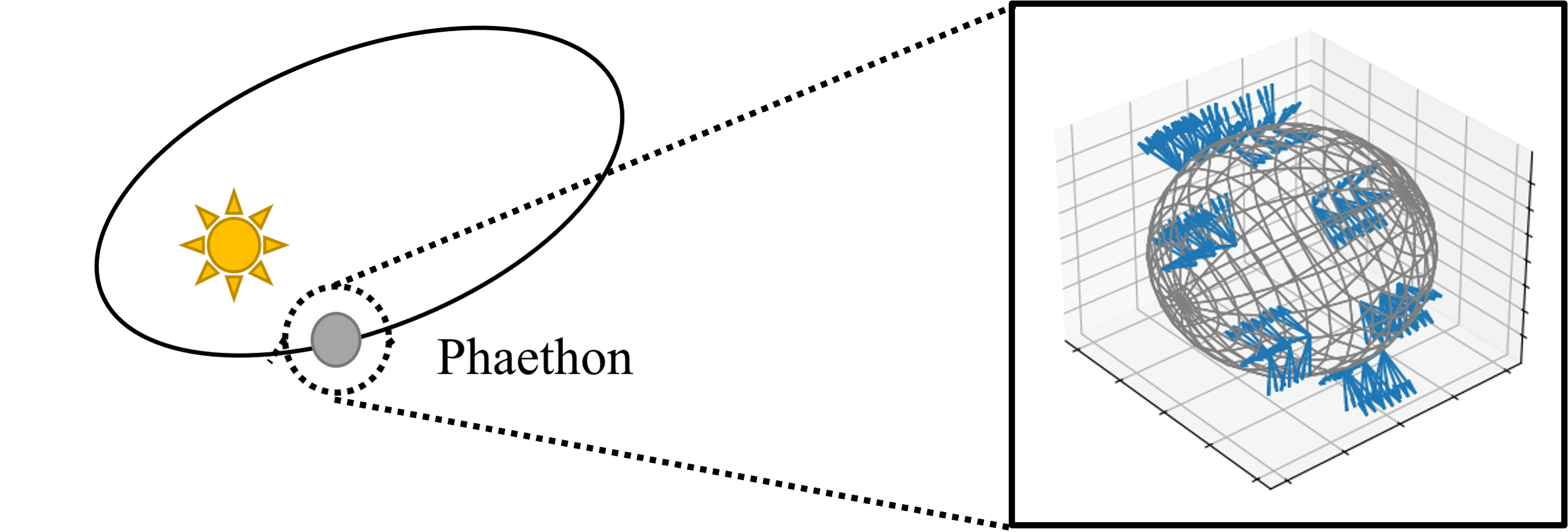}}
 \caption{Schematic illustration of dust ejection in our model. In the figure, the gray sphere corresponds to the Hill radius of Phaethon ($r=78$ km from Phaethon's center of mass at present-day orbit). The dust particles have small outward terminal velocities (the blue arrows) on the Hill sphere ($v_r=1$~m s\textsuperscript{-1}).
 \label{fig:initialsphere}}
\end{figure}

The ejection of a particle can be described by its position and velocity vector ($r$, $\theta$, $\phi$, $v_{r}$, $v_{\theta}$, $v_{\phi}$), each corresponding to the radial distance, polar and azimuthal angle, respectively, in the Phaethon-centric frame. Ejection velocity by rotational instability could be assumed to be near the escape speed of the parent body. However, the actual ejection velocity can be higher, depending on its cohesive strength \citep{2014ApJ...789L..12H}. Notably, P/2013 R3 is thought to have been broken up by rotational instability, and its fragments had a velocity dispersion of $0.33 \pm 0.03$~m/s, which is more than 50\% higher than the assumed escape speed, $\sim 0.2$ m/s, of the progenitor \citep{2017AJ....153..223J}. Taking this into account, we assumed $v_r = 1$~m s\textsuperscript{-1} and $r$ as the Hill radius of Phaethon at perihelion:

\begin{equation}
r_\mathrm{H} \approx a (1-e) \left(\frac{M}{M_{\odot}}\right)^{1/3} ,
\label{eq:3}
\end{equation}
\noindent where Phaethon's mass $M=1.16 \times 10^{17}$ g is calculated assuming a spherical body with its mass density and diameter reported in \citet{2018A&A...620L...8H}. With this initial setting, we assumed that the particle had an initial surface velocity that was about 1.4 times higher than the escape velocity but had lost most of its kinetic energy when the particle reached the Hill radius (see Fig. \ref{fig:initialsphere}).

Focusing on rotational instability as the cause of the dust ejection, we further assume that the particles are ejected in the lower latitude region of Phaethon and follow its rotational direction. As such, we binned the remaining vector components ($\theta$, $\phi$, $v_{\theta}$, $v_{\phi}$) with 20 bins in $[-\pi/18, \pi/18], [0,2\pi], [-\pi/18, \pi/18], [0,\pi/2]$ respectively, resulting in 320\,000 particles per simulation. The vectors were then rotated to match Phaethon's current rotational axis \citep{2016A&A...592A..34H}. This initialization process is visualized in Fig. \ref{fig:initialsphere}. We then combined the position and velocity vectors of the dust particles with the orbital vectors of Phaethon at the ejection epoch. All particles are considered massless in the simulation. Therefore, all particles are subject to external gravity and radiation forces but do not exert gravity. The particles were then forward integrated until the present day (JD 2459000.5) along with the 8 planets and Phaethon (to take into account the initial gravitational influence from the parent body) with an output interval of 500 days.

\section{Results} \label{sec:result}

In this section, we first scrutinize the simulation results under the MOID condition and then compare them with the Geminid observations in terms of solar longitude. 

\subsection{Stream location based on minimum orbital intersecting distance} \label{subsec:clone}

\begin{figure}
  \resizebox{\hsize}{!}{\includegraphics{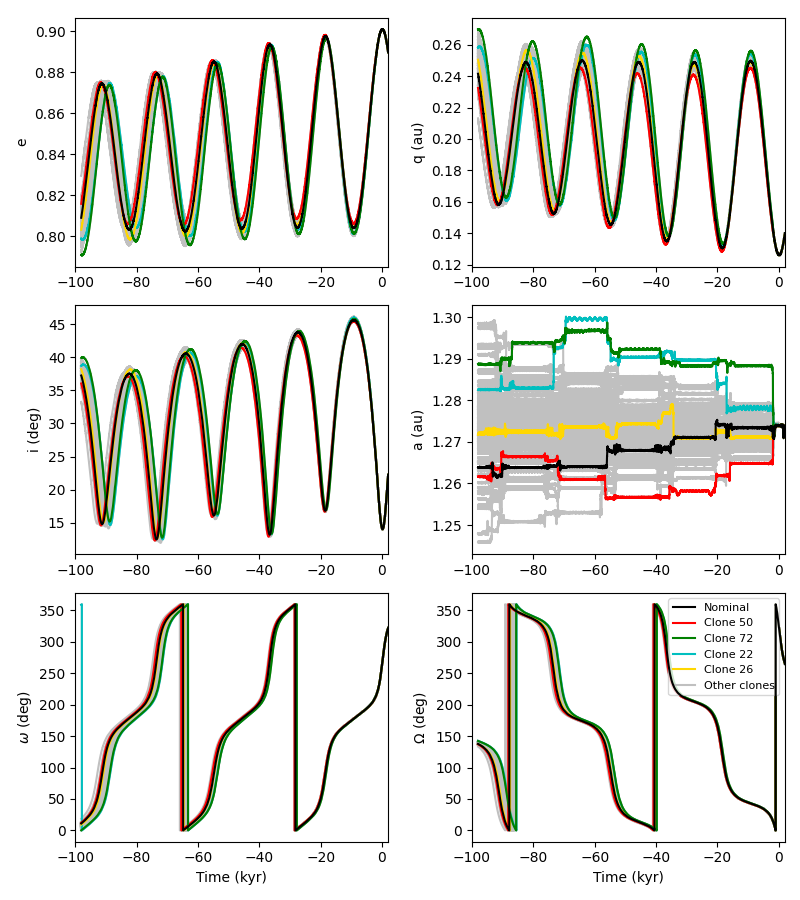}}
  \caption{Orbital history of 100 Phaethon clones from backward integration. The colored lines are clones we used for the dust sources.}
    \label{fig:backward}
\end{figure}
Fig. \ref{fig:backward} shows the orbital history of Phaethon from backward integration. It is important to note that the deviation between the orbital elements of nominal Phaethon and its clones noticeably increased $\sim 4000$ years ago, triggered by planetary perturbations \citep{2016A&A...592A..34H}. To eliminate this uncertainty in Phaethon's orbital history, we selected 2 "extreme" clones that correspond to the maximum and minimum semimajor axis after the first 10000 years of backward integration. We then randomly chose two additional clones. The initial orbital elements of these clones are given in Table \ref{table:clone}, and their backward integration results are also marked with colored lines in Fig. \ref{fig:backward}.

\begin{figure}
\resizebox{\hsize}{!}{\includegraphics{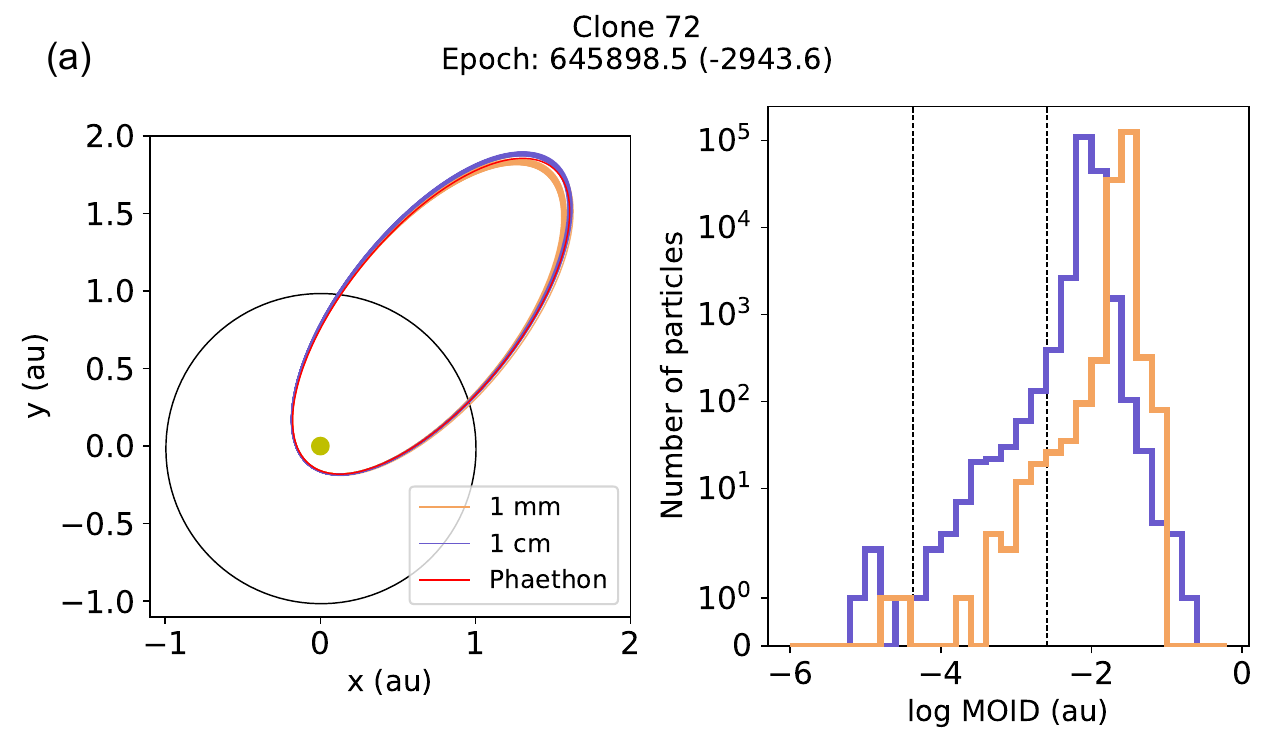}}
\resizebox{\hsize}{!}{\includegraphics{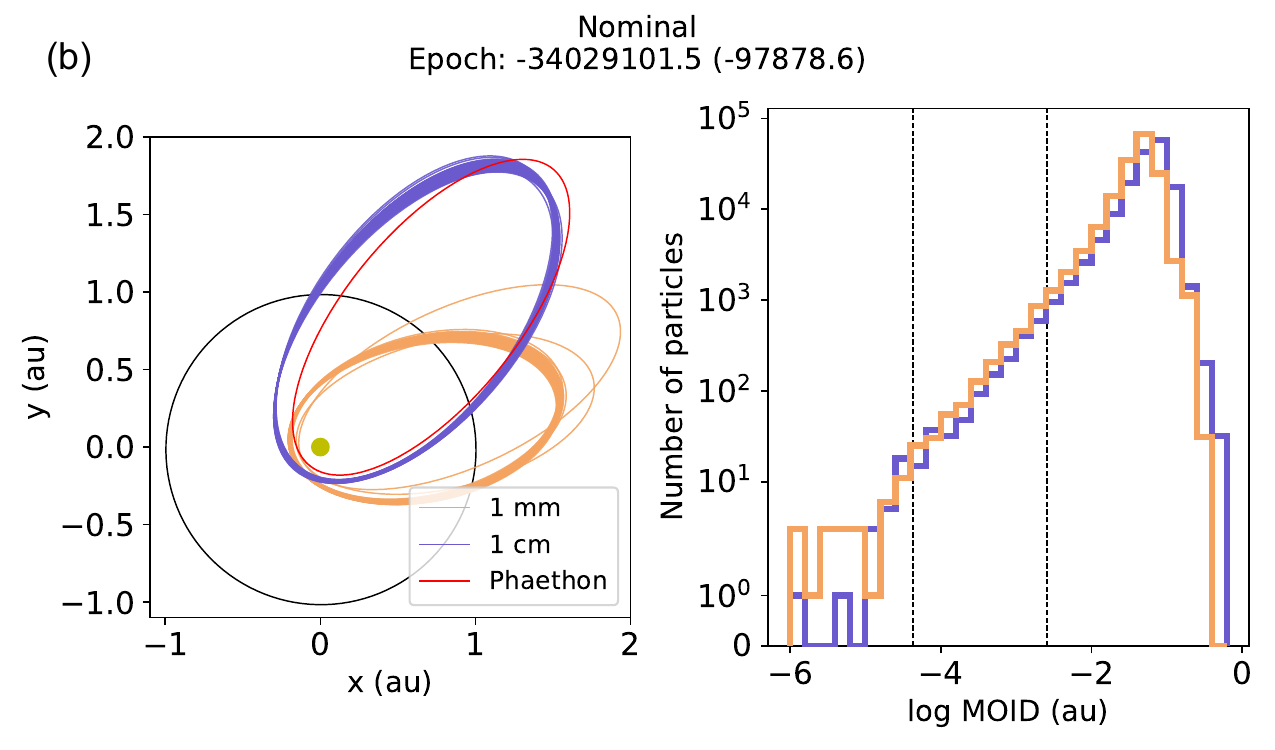}}
\caption{Examples of the dust simulation result from nominal Phaethon ejection. The ejection epoch is labeled at the top in Julian date (and its corresponding year in bracket). The left plots show random samples of particle orbits on the ecliptic plane at the present day (JD 2459000.5). The right plots are the histograms of the MOID of particles to Earth, with the Earth radius and Earth-Moon distance marked by dotted lines.}
 \label{fig:nominalhist}
\end{figure}

To determine the existence of particles that interact with Earth, we employed the minimum orbit intersecting distance (MOID). The idea is that each simulation particle represents numerous dust particles sharing the same orbital elements but a random mean anomaly. Therefore, even if the simulation particle does not directly contact Earth if its MOID is small enough, then there will be a corresponding dust particle in reality that collides with Earth. Similar approaches to identify interactions between Earth and meteors have been used in studies such as \citet{2021JGRE..12606817M} and \citet{2021MNRAS.507.4481R}. We set two MOID criteria: Earth's radius ($\sim4.26 \times 10^{-5}$ au) and the semi-major axis of the lunar orbit ($\sim2.57 \times 10^{-3}$ au) because the meteor observations have been conducted either from Earth surface or the lunar orbit. We calculated the MOIDs of the particles using the code by \citet{2013AcA....63..293W}. For example, in Fig. \ref{fig:nominalhist} (a), we found particle MOIDs smaller than the lunar orbit and even Earth's radius, indicating the presence of particles interacting with the Moon and Earth in this case. It is worth noting that the stream's core locates at $\sim 0.01$ au from Earth at its closest, implying that Earth is merely grazing the stream. This evidence implies that if a distance condition at a scale of 0.01 au is used, one risks over-representing particles from the stream core. In other words, using lenient distance conditions of $\ge 0.01$ au does not accurately represent Geminid meteoroids. In the right column of Fig. \ref{fig:nominalhist}, we show examples of MOID distribution of the particles from our dynamical simulation. However, while MOID is a strong indicator of how close the dust approaches Earth, where the dust crosses is also another important condition we should consider. 
 
\begin{figure}
 \resizebox{\hsize}{!}{\includegraphics{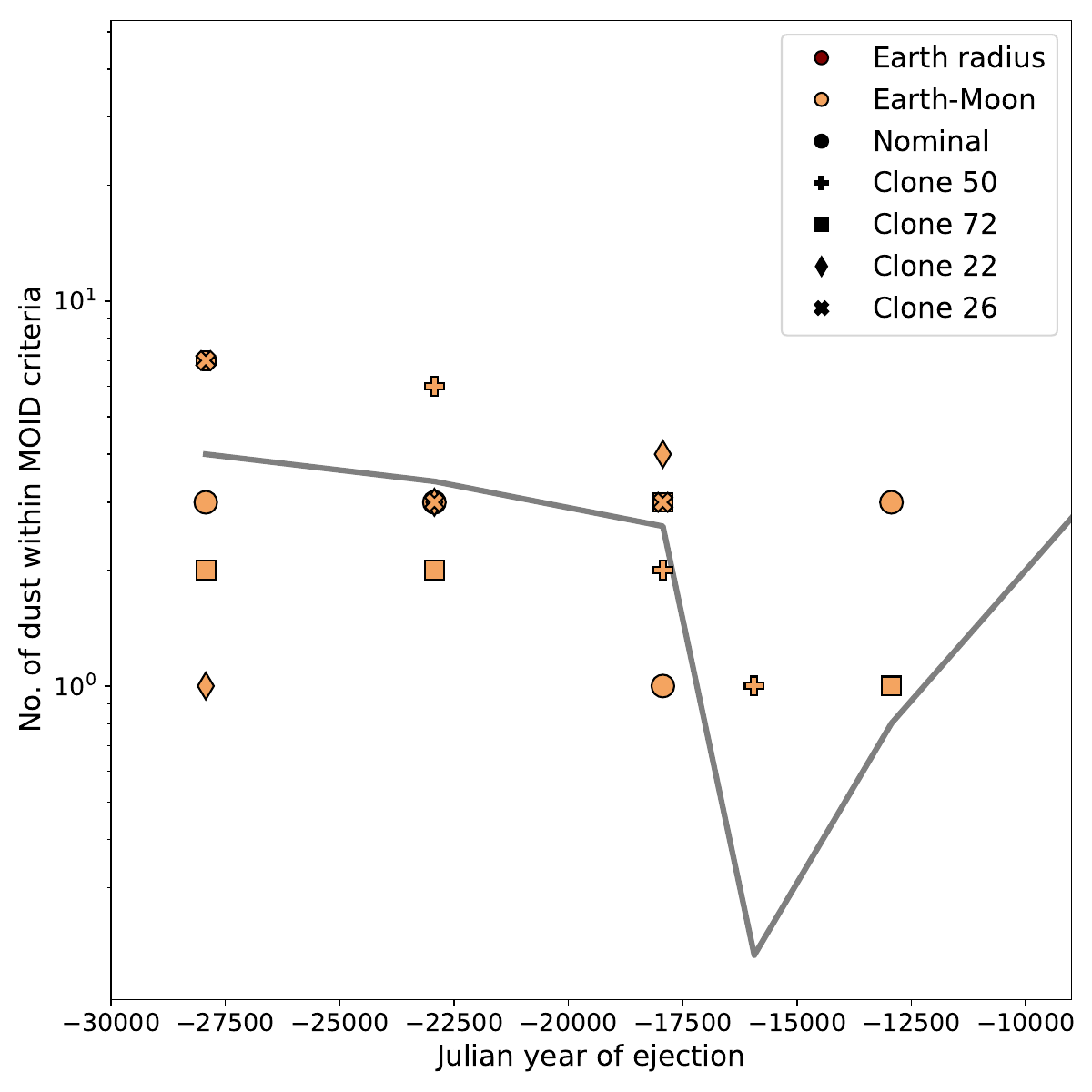}}
 \resizebox{\hsize}{!}{\includegraphics{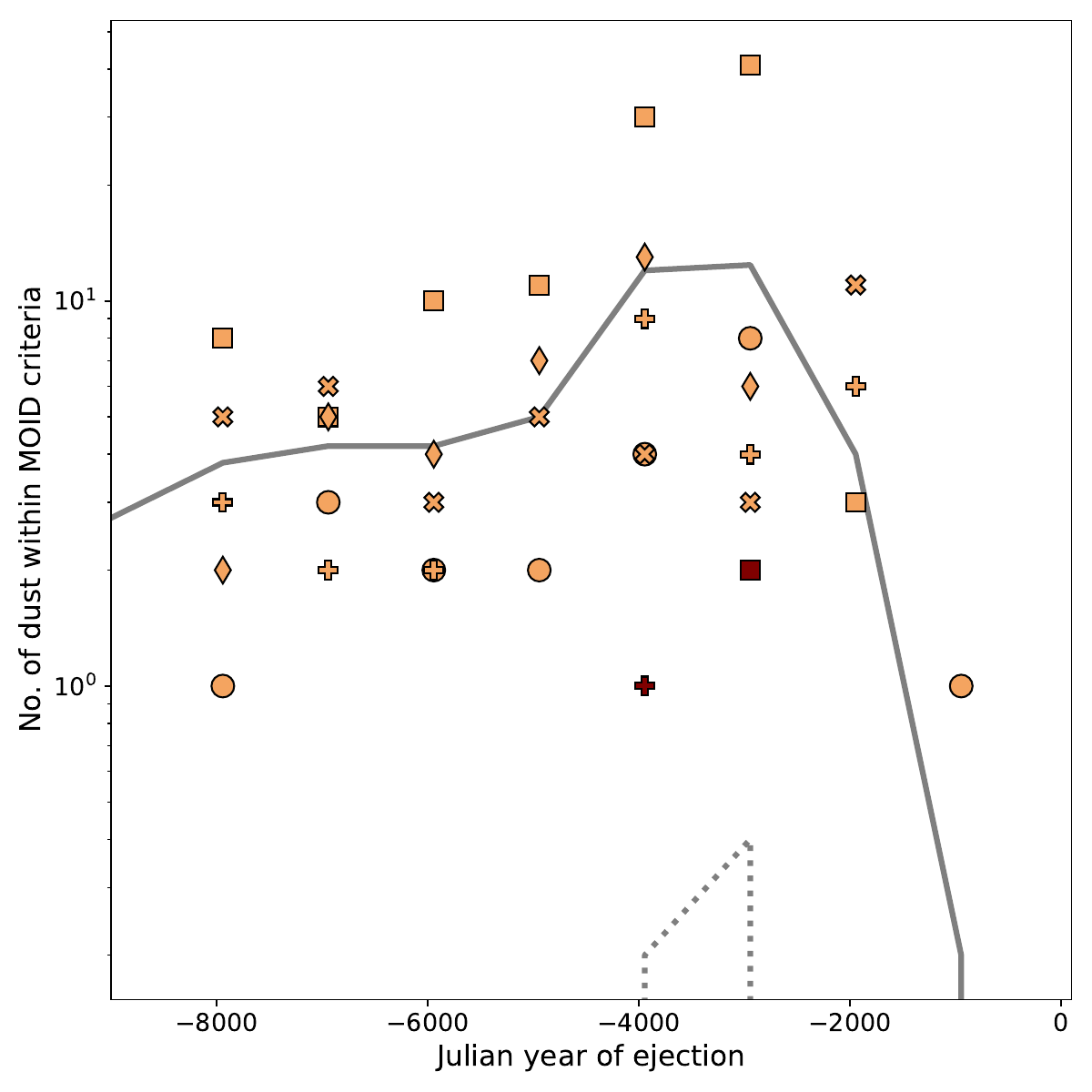}}
\caption{Number of 1-mm particles satisfying the MOID criteria from each ejection epoch with different ranges of ejection epoch (top: 10000 - 30000 years ago, bottom: 2000 - 10000 years ago). Different colors represent the MOID criteria of Earth radius (darker color) and Earth-Moon distance (lighter color), respectively. The dotted line shows the mean-averaged values of the number of particles with MOIDs within Earth's radius from nominal and clone parent bodies. The solid line is the equivalent of the dotted line for the Earth-Moon distance criterion. Simulation results with no particles within the MOID criteria are not present, but was included when calculating the mean average.}
 \label{fig:moidnum1mm}
\end{figure}

\begin{figure}
 \resizebox{\hsize}{!}{\includegraphics{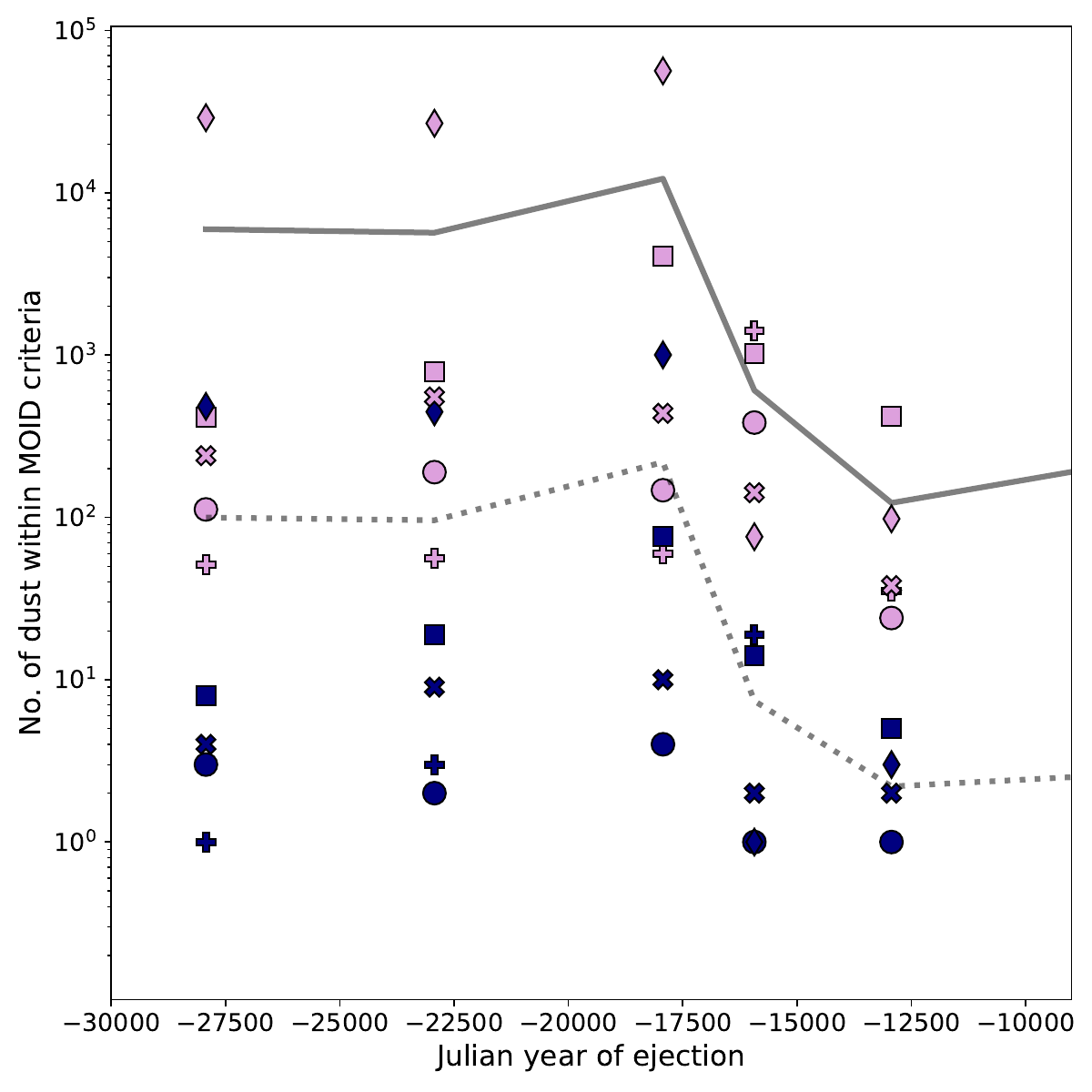}}
 \resizebox{\hsize}{!}{\includegraphics{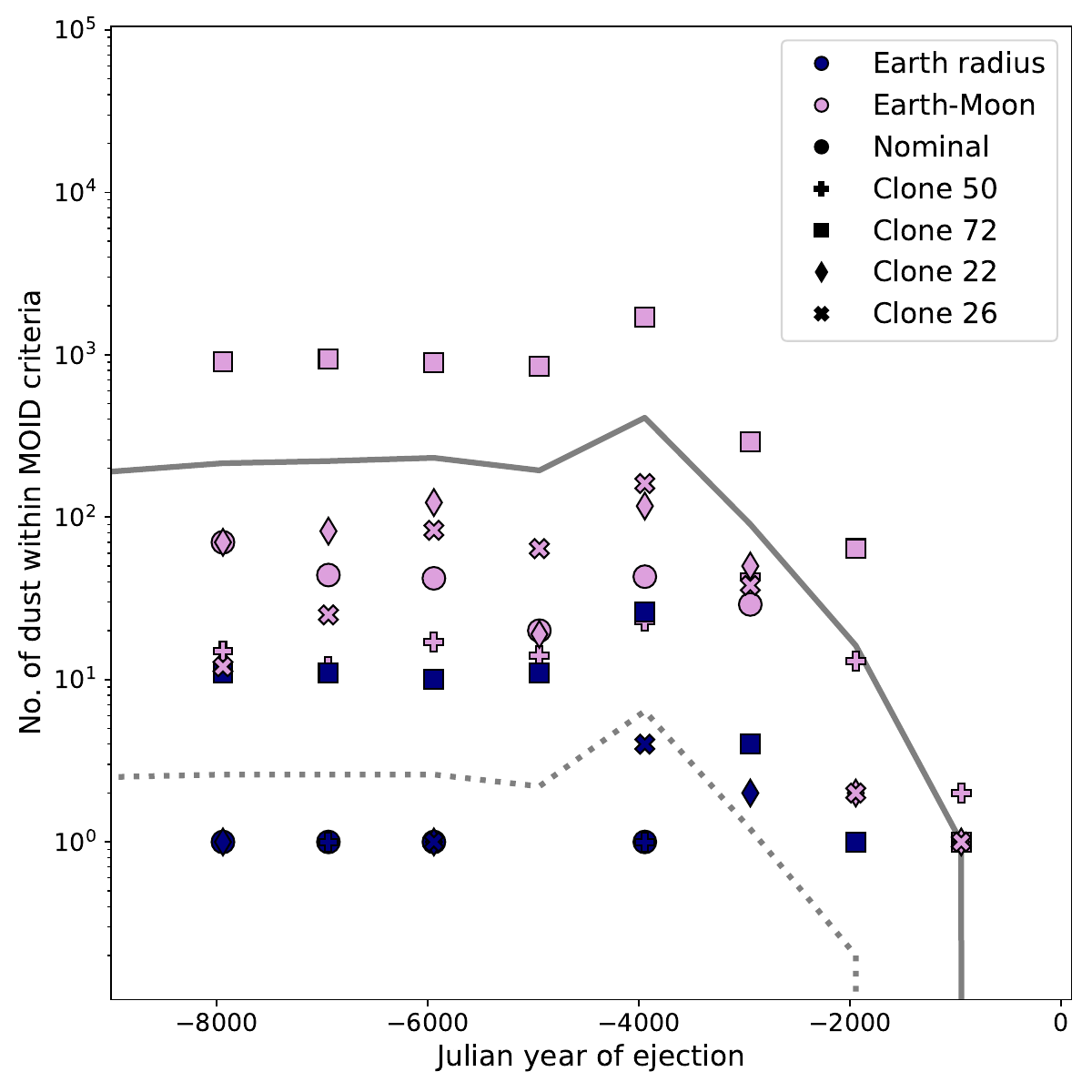}}
\caption{Same as Fig. \ref{fig:moidnum1mm}, but for 1 cm particles}
 \label{fig:moidnum1cm}
\end{figure}

Figs. \ref{fig:moidnum1mm} and \ref{fig:moidnum1cm} summarize the number of interacting particles from each ejection epoch for 1-mm and 1-cm particles. It is found that the interacting particles significantly reduced if the particles were ejected after 4000 years ago. This can be explained by the fact that the current Phaethon's orbit is far from the Earth, and apart from occasional weak gravitational perturbations of terrestrial planets, there is not enough external force that can transport the particles from Phaethon's orbit to Earth-interacting orbit in time. Further back in time, we found Phaethon's orbital uncertainty increases, resulting in larger variations in the number of interacting particles between the nominal and clone cases. We would also like to point out that the number of 1-mm particles are much smaller compared to that of 1-cm particles. Nonetheless, for all ejection epochs, our simulation produced Earth-dust interactions from the initial Phaethon ejections in both the nominal and clone cases. 

\subsection{Stream location based on solar longitude} \label{subsec:streamloc}

Although the MOID criterion is one of the strong indicators of the stream--Earth interaction, it does not necessarily mean that the simulation results exactly reproduce the spatial distribution of Geminids. For example, Fig. \ref{fig:nominalhist} (b) shows a stream from an ejection 100\,000  years ago, and one can visibly notice that the stream has shifted considerably from Phaethon's orbit at present day. Indeed, although a large number of particles in this stream interacts with Earth, the intersection is too far off compared to observed Geminid locations. As a result, we can henceforth completely reject the models from 100\,000 years ago. For stream models from other ejection epochs, this intersection discrepancy can be less obvious. Thus, it is essential to introduce an additional criterion to select a suitable Geminid-like model stream.

\begin{figure}
\resizebox{\hsize}{!}{\includegraphics{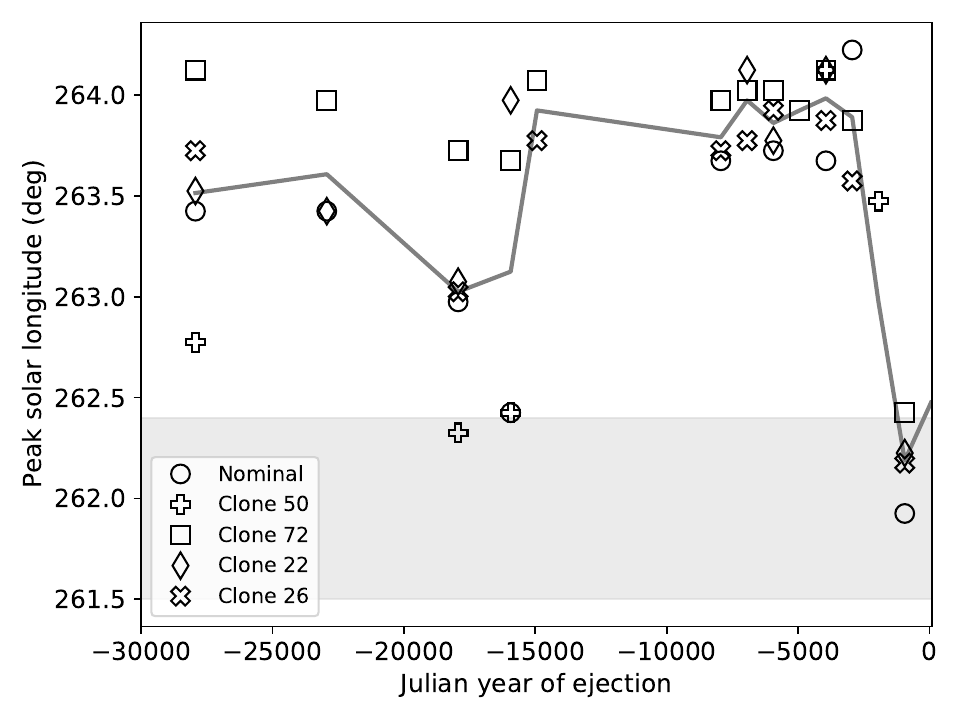}}
\caption{Mode solar longitude of 1 cm particles within the lunar distance MOID criterion. If there was more than one maximum peak, we registered the mean solar longitude of the peaks as the corresponding data point. The blue line is the mean of the data points. The shaded area marks the maximum activity plateau ($261.5$ - $262.4$\degr) reported by \citet{2004EM&P...95...27R}.
\label{fig:slpeak}}
\end{figure}

Most well-observed Geminid showers have their peaks at the solar longitude of $\sim 262$\degr \citep{2004EM&P...95...27R, 2006MNRAS.367.1721A,2009CoSka..39....5Z,2019JIMO...47..180R}. More specifically, \citet{2004EM&P...95...27R} reported a maximum activity plateau at 261.5\degr - 262.4\degr that is consistent for 60 years of observations. Fig. \ref{fig:slpeak} shows the mode solar longitude, in other words, the solar longitude corresponding to the peak activity of 1-cm particles from each model. As mentioned in Sect. \ref{subsec:clone}, due to the lack of 1-mm particles relative to 1-cm particles within the MOID criteria, extracting the mode solar longitude from 1-mm particles is statistically unreliable and thus was ignored (we will address this issue in Sect. \ref{subsec:collisional}). We found that several models at ejection epochs of 3000, 18\,000, and 20\,000 yrs ago are marginally in the range of the maximum activity plateau, although most of our simulation results have a systematical offset of the solar longitude by $\sim 1$\degr, as noticed in previous studies \citep{2007MNRAS.375.1371R,2016MNRAS.456...78R}. We will discuss these cases in depth in Sect. \ref{subsec:activity}.

\section{Discussion} \label{sec:discussion}

\subsection{Replenishment of mm-sized particles from collisional cascading} \label{subsec:collisional}

As mentioned in Sect. \ref{subsec:streamloc}, the simulation result shows a relative shortage of 1-mm particles within the MOID criteria. Fig. \ref{fig:moidnum1mm} indicates that the paucity of 1-mm particles becomes more severe in older epochs. Over a timescale of $>10^4$ yr, 1-mm particles deviate significantly from the orbital trend of 1-cm particles and even Phaethon due to a complex interplay of the Poynting-Robertson drag \citep{1979Icar...40....1B}, Kozai-Lidov mechanism \citep{1962P&SS....9..719L,1962AJ.....67..591K}, and perihelion precession. The Poynting-Robertson drag reduces the semimajor axis and eccentricity of the particles, thereby affecting the gravitational influence of the planets in the three-body relation with the Sun and a test particle and the resulting Kozai-Lidov oscillation and perihelion precession. Thus, particles smaller than $\sim 1$-mm should be excluded from the Geminid stream.

Our simulation results would be inconsistent with observations because particles as large as 1 mm have been observed as Geminids meteor showers. Such small particles ($\sim1$ mm) may have been supplied at a different epoch from the larger particle ejection. However, this hypothesis is unlikely because particles ejected recently (within the past 2000 years. see, Fig. 4) have no dynamical path to reach the near-Earth region due to weak gravitational perturbations by the terrestrial planets.

Another potential solution is that the mm-sized particles were produced through the collisional cascading of larger particles. Early calculations such as \citet{1982mmis.proc..184T,1986MNRAS.218..185S} estimated the lifetime of the Geminids based on the collision probability with interplanetary particles, assuming various zodiacal cloud distribution models, and derived a range of $\sim 10^5$ yr for 1 cm particles. On the other hand, the recent Interplanetary Meteoroid Environment Model 2 (IMEM2, \citealt{2019A&A...628A.109S}), which is an interplanetary dust cloud model derived from dust ejected by Jupiter-family comets, Halley-type comets, and main-belt asteroids, showed that in the centimeter regime, the collisional lifetime of a dust particle moving along Phaethon's orbit can range from $\sim10^3 - 10^5$ yr. Meanwhile, the lifetime of mm-sized particles does not exceed $10^3$ yr. Therefore, a significant portion of <1-mm particles in the current Geminids was likely produced through the collisional cascading of >1-cm particles in $\sim 10^4$ yr. In this case, we can assume that the more recently created mm-sized particles would closely follow the orbits of cm-sized particles, and thus, the results from cm-sized particles are applicable to the mm-scale as well. However, it is worth noting that the observational traits of sporadic meteors do not agree with the various dynamical models of the zodiacal cloud, including IMEM2. In order to compensate for this discrepancy, it was found that particles of >0.1 mm should have a longer lifetime than calculated in the dynamical models \citep{2011ApJ...743..129N,2014ApJ...789...25P}. As a result, it is not uncommon to use an artificial factor, denoted as $F_\mathrm{coll}$ in \citet{2014ApJ...789...25P} and \citet{2019A&A...628A.109S}, to increase the lifetime of larger particles in the zodiacal cloud by one order of magnitude or more. Then, one might question if the above-mentioned lifetimes of mm- and cm-sized Geminid particles are greatly underestimated and thus, their collisional effects are negligible. However, since Phaethon, a near-Earth object, is not part of the dust source considered in IMEM2, we suggest that the Geminids are not subject to the artificial factor, which only accounts for the collisions between the dust cloud inhabitants.

\subsection{Estimate of the Geminid stream mass} 
\label{subsec:massestimate}

Although we have demonstrated that particles ejected with low velocity can form a meteoroid stream, it is important to investigate the validity of the number of Earth-interacting particles. This analysis makes two assumptions: (1) the interacting particles represent the Geminid mass influx on Earth, and (2) the stream comprises particles originating from the ejection epoch under consideration. Based on these assumptions, we estimated the total mass of the stream and compared the mass with known estimates. 

Our mass estimation process is as follows. In general, the mass distribution of a dust cloud can be given by:

\begin{equation}
dN(m) = k ~\left(\frac{m}{m_0}\right)^{-s} dm~,
\label{eq:dN}
\end{equation}

\noindent or

\begin{equation}
n(m) = k' ~\left(\frac{m}{m_0}\right)^{-s} ~,
\label{eq:N}
\end{equation}

\noindent where $N(m)$ is the cumulative mass distribution of the infalling Geminids with mass larger than $m$. We convert this to $n(m)$, which is the number of dust particles with mass $m$. Note that $N(m)$ and $n(m)$ do not characterize the entire Geminid stream but a portion that interacts with the Earth's orbit. $k$ and $k'$ are constants that characterize the stream mass for reference mass $m_0$, which we arbitrarily choose to be 1 g.

Because the dust population in our simulation is not continuous in mass, we modified Eq. (\ref{eq:N}) as below:

\begin{equation}
\begin{split}
M_\mathrm{C,tot} & = \sum_{m} M_\mathrm{C}(m)\\ 
 & = \sum_{m} n(m) \Delta m\\
 & = \sum_{m} k' ~\left(\frac{m}{m_0}\right)^{-s} \Delta m.
\end{split}
\label{eq:4}
\end{equation}

\noindent $m$ is derived by assuming that they are spherical particles of 2.9 g/cm\textsuperscript{3} with radii of 0.1 mm, 1 mm, 1 cm, and 10 cm, which roughly corresponds to the mass range used in \citet{2017P&SS..143...83B} to estimate $M_\mathrm{C,tot}$. $\Delta m$ are logarithmically equally spaced intervals centered around 0.1 mm, 1 mm, 1 cm and 10 cm. The differential mass distribution power index from the meteor and lunar impact monitoring was reported to be $s =1.68 \pm 0.04$, while the observed mass influx was $M_\mathrm{C,tot}$=$1.5 \times 10^{8}$ g \citep{2017P&SS..143...83B}. However, our simulation indicates a total stream width, or shower duration, of only ~3 days, which is much shorter than the 14-day activity profile derived in \citet{2017P&SS..143...83B}. Therefore, we only considered the activity of 3 days around the peak. We calculated that this portion of the profile represented 0.49 of the total area, resulting in $M_\mathrm{C,tot} = 7.35 \times 10^{7}$ g.  Finally, we can then calculate the constant $k'$ and consequently compute the observed mass influx for particles of mass $m$, $M_\mathrm{C}(m)$. We note that the power index $s$ used in this calculation characterizes the dust population that falls into Earth, but not necessarily the entire Geminid stream. Nonetheless, we point out that the use of this $s$-value in this section is only for the purpose of calculating $k'$ by retracing the original calculations of \citet{2017P&SS..143...83B}.

We utilized a variation of the mass estimation formula proposed by \citet{1989MNRAS.240...73H}. The original equation assumes a circular cross-section for the meteoroid stream and a uniform meteoroid flux throughout this cross-section, with the flux equivalent to the maximum shower activity. Under these conditions, the authors defined $M_\mathrm{F}$ as the meteoroid flux integrated over the cross-section, deriving it with the following relation: 

\begin{equation}
M_\mathrm{F} = \frac{M_\mathrm{C,tot}}{t} \frac{A V_\mathrm{H}}{\pi R_\mathrm{E}^2 V_\mathrm{G}}~,
\label{eq:5}
\end{equation}

\noindent where $t$ is the shower duration, which is the time it takes for Earth to pass through the stream. $V_\mathrm{H}$ and $V_\mathrm{G}$ are the heliocentric and geocentric stream velocities, respectively. $A$ is the cross-section area of the stream, while $R_\mathrm{E}$ and $V_\mathrm{E}$ represent the Earth's radius and orbital velocity. The factor, $\pi R_\mathrm{E}^2 V_\mathrm{G}$, can be interpreted as the volume of the tube formed by Earth's movement relative to the stream during $t$. Similarly, $A V_\mathrm{H}$ is the total volume that the stream passes through during $t$. Thus, the rightmost fraction in Eq. (\ref{eq:5}) is the ratio between the volume of the stream passing through a perpendicular plane and the volume created by the passing of Earth within the stream as they intersect each other. Following their initial assumptions, \citet{1989MNRAS.240...73H} equated this ratio of volumes to the ratio of meteoroid numbers and subsequently managed to extrapolate observed Geminid mass influx into the total Geminid stream mass. In this work, we are able to count the number of particles that interact with Earth and those that do not, removing the need for such assumptions. Therefore, Eq. (\ref{eq:5}) can be changed to:

\begin{equation}
M_\mathrm{F}(m) = \frac{M_\mathrm{C}(m)}{t} \frac{N_\mathrm{tot}(m) V_\mathrm{H}}{N_\mathrm{E}(m) V_\mathrm{G}},
\label{eq:6}
\end{equation}

\noindent where $N_\mathrm{E}(m)$ and $N_\mathrm{tot}(m)$ are the number of simulation particles interacting with Earth, or in other words, satisfying the MOID criteria, and the total number of simulation particles of mass $m$, respectively. We used $t=3$ days, which is the approximate duration of the shower based on our simulation. Using the mean orbital period $P$ of the simulation particles, the total stream mass can then be calculated as:

\begin{equation}
M_\mathrm{tot} = \sum_{m} M_\mathrm{F}(m) P(m).
\label{eq:7}
\end{equation}

Although the original $M_\mathrm{C}$ value was derived assuming a mass range of 0.1 mm - 10 cm, we were computationally limited to performing simulations for the entire mass range. The Poynting-Robertson timescale for 1 cm and 10 cm particles are $\sim1 - 10$ Myr. Given that our simulation timescale spans $\sim 10\,000$ years or less, the radiative effects on the orbital motion of these particles are negligible. Hence, it is reasonable to extrapolate the simulation results of 1 cm particles to those of 10 cm particles. However, the same cannot be assumed for 0.1 mm, whose Poynting-Robertson timescale is $\sim 10\,000$ yr. To overcome this, we rely on the mass population distribution power-law index to derive the mass of 0.1-mm particles from our more statistically reliable 1-cm particle population. The mass index $s$ used in Eq. (\ref{eq:dN})-(\ref{eq:4}) describes the infalling Geminid population, but not necessarily the entire Geminid population, as the vast majority of the Geminid stream does not interact with Earth. 

In the scenario of rotational instability, it is reasonable to hypothesize that the resulting dust population would retain the mass distribution index of its regolith phase. With this in mind, we employ two mass indices obtained by the Hayabusa2 mission on asteroid Ryugu. The first is the global boulder cumulative size distribution index of 2.65 \citep{2019Icar..331..179M}. The second is the cumulative power-law index of 3.88 for the cumulative size distribution of Ryugu regolith samples \citep{2022NatAs...6..214Y}. It is worth noting that the larger size index of the returned regolith sample is attributed to the fragmentation caused by the impact experiment. If we convert the cumulative size distribution index to the differential mass distribution index, the values are 1.88 and 2.29 respectively. In comparison, this value is slightly higher than the observed index found by \citet{2017P&SS..143...83B} ($1.68 \pm 0.04$) and close to the index used by \citet{2017P&SS..143..125R} (2.17-2.438). With this considered, the 0.1 mm particles will have approximately $10^{3.76}$ or $10^{1.3}$ times the total mass of 1 cm particles, depending on the index. In several models, we had models with no 1-mm particles satisfying the MOID criteria, leading to a divided-by-zero problem in Eq. \ref{eq:6}. As such, if no 1-mm particles are present within the MOID criteria, we approximate the mass of 1 mm particles as $10^{1.88}$ or $10^{0.65}$ times that of 1 cm particles, similar to our approach with 0.1 mm particles.

\begin{figure}
 \resizebox{\hsize}{!}{\includegraphics{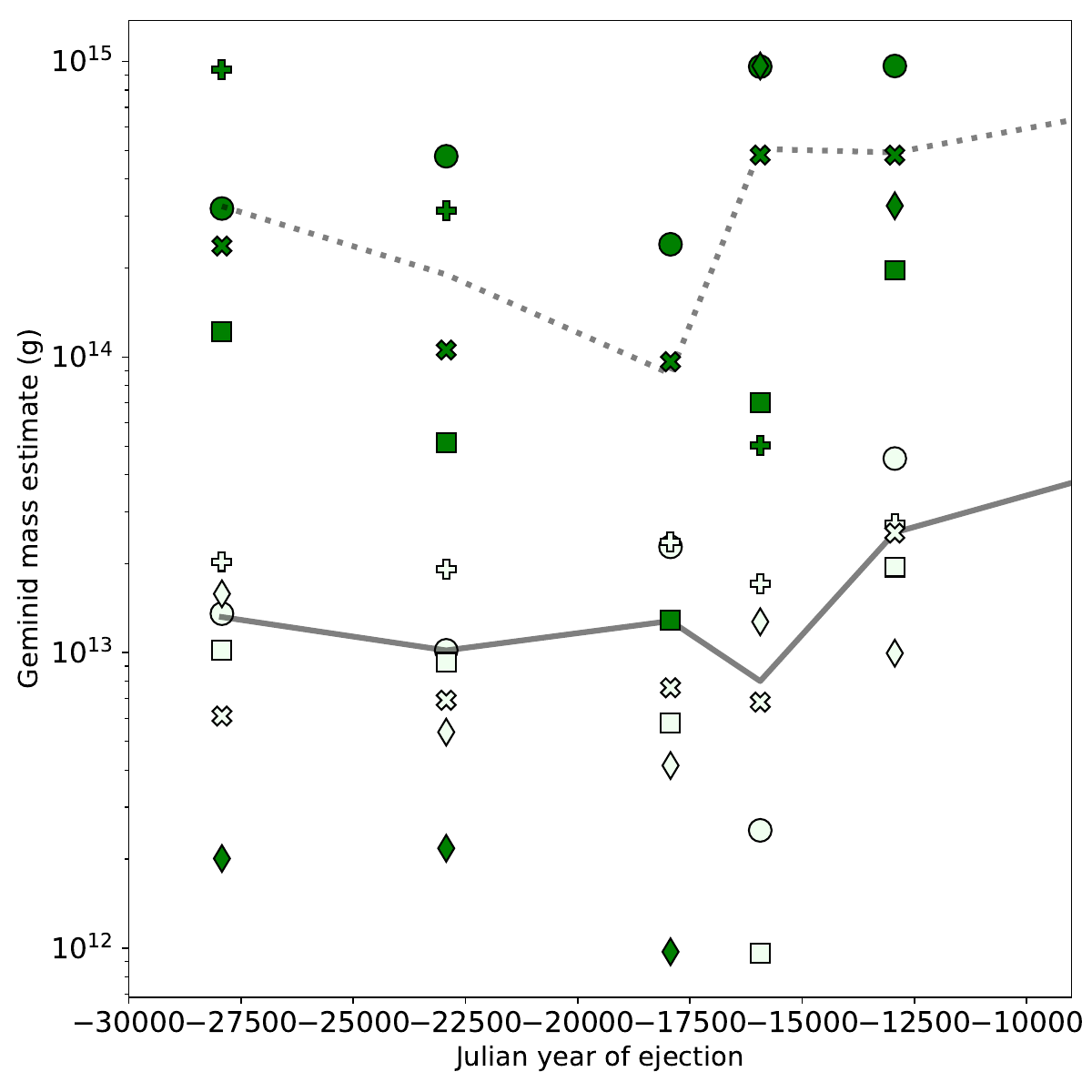}}
 \resizebox{\hsize}{!}{\includegraphics{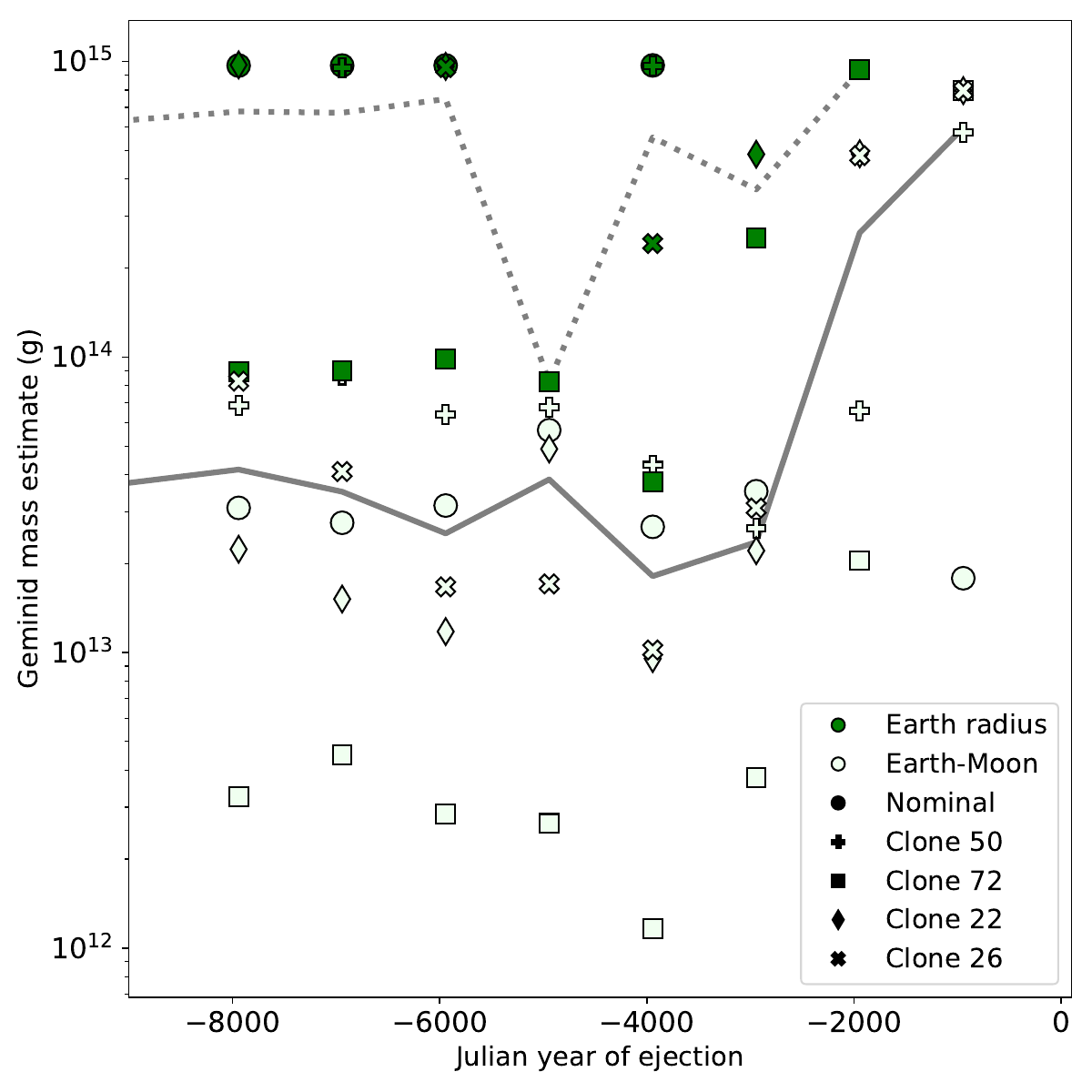}}
\caption{Mass estimate of the Geminid stream for all the simulations performed in this study. The index by \citet{2022NatAs...6..214Y} is assumed. Top panel displays data points ranging from $\sim 15 - 30$ kyr ago, while the bottom panel shows the values from $\sim 2 - 10$ kyr ago. The letters A and B mark the optimal models chosen in Sect. \ref{subsec:activity}.
\label{fig:moidmass}}
\end{figure}

\begin{figure}
 \resizebox{\hsize}{!}{\includegraphics{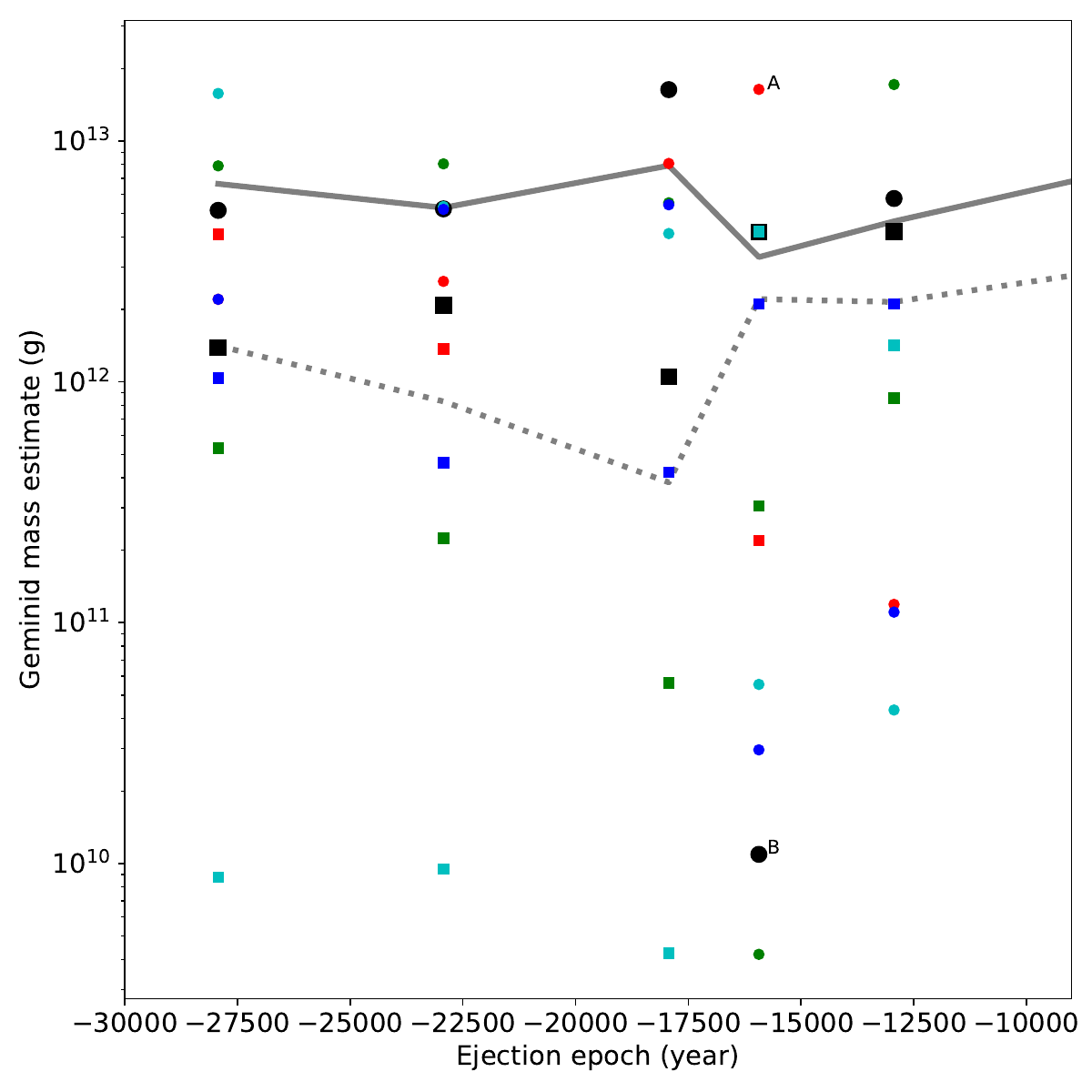}}
 \resizebox{\hsize}{!}{\includegraphics{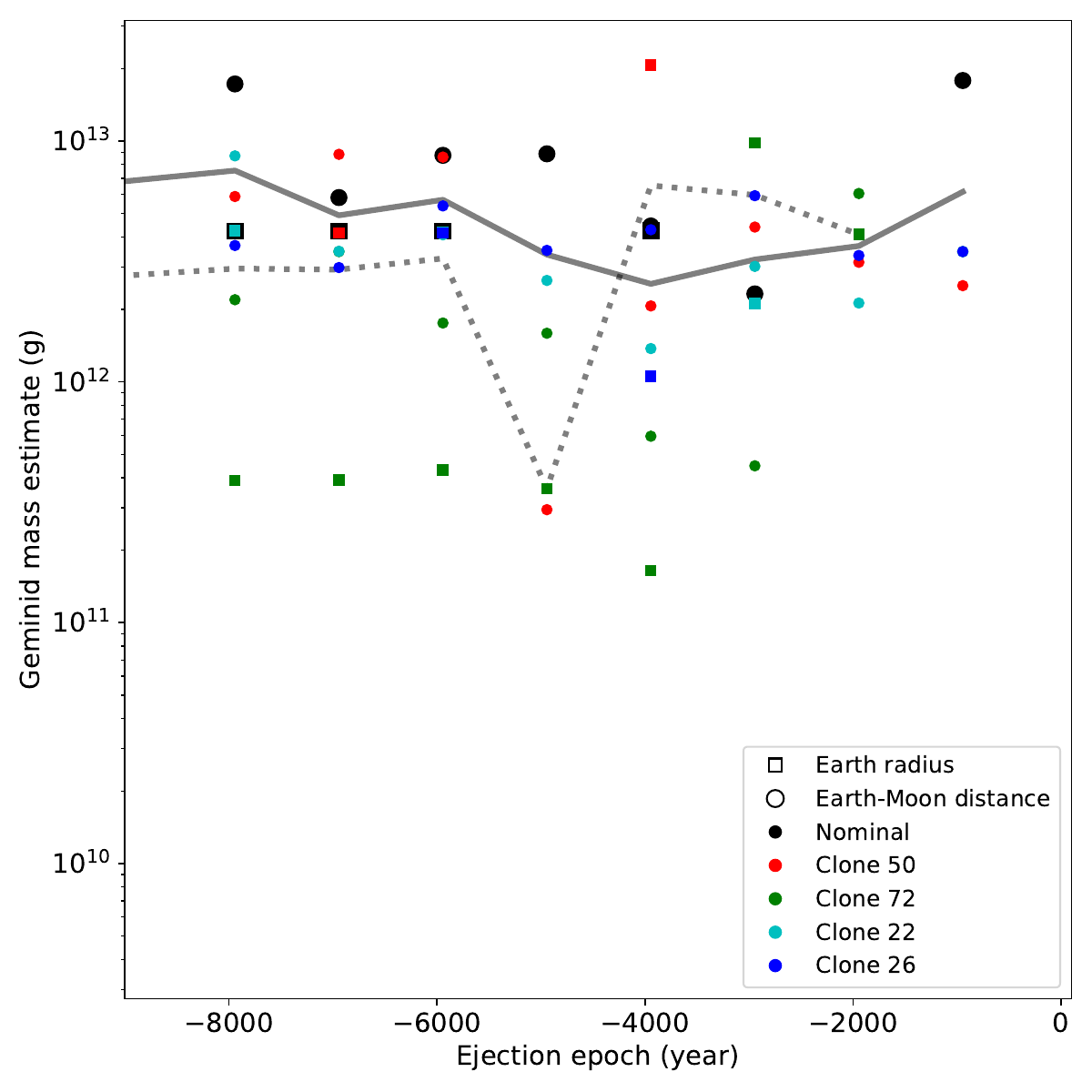}}
\caption{Same as Fig. \ref{fig:moidmass}, but for the index by \citet{2019Icar..331..179M}.
\label{fig:moidmass2}}
\end{figure}

Figs. \ref{fig:moidmass} and \ref{fig:moidmass2} illustrate the estimated mass of the dust streams from all models. Our calculation of the total stream mass yields a range of $\sim10^{10} - 10^{14}$ g. Compared to Phaethon's mass, which is currently estimated at around $10^{17}$ g \citep{2018A&A...620L...8H}, the upper bound corresponds about 0.1\%, or 42 cm of its surface layer. Interestingly, this range overlaps with the mass range of the Phaethon dust trail, $\sim 10^{13} - 10^{15}$ g, found by the Wide-field Imager on the Parker Solar Probe under the assumption that the trail consists of 0.5 mm particles \citep{2022ApJ...936...81B}.

Meanwhile, past estimations of the Geminid stream mass range widely depending on the author and methodology, such as $1.6 \times 10^{16}$ g \citep{1989MNRAS.240...73H,2017P&SS..143...83B}, $(1.4 \pm 0.5) \times 10^{15}$ g \citep{1994A&A...287..990J} and $5.0 \times 10^{15} - 4.3 \times 10^{17}$ g \citep{2017P&SS..143..125R}. Our model does not replicate the total width of the Geminid activity profile (we will discuss this more in Sect. \ref{sec:limits}) and therefore, the derived mass estimate could be underestimated by a factor of 3-5. Nonetheless, the difference of orders of magnitudes between our result and the estimates from previous studies should be addressed. In the following paragraphs, we will provide the reasoning for this disparity. 

In the cases of \citet{1989MNRAS.240...73H,1994A&A...287..990J,2017P&SS..143...83B}, the calculation is based on Eq. (\ref{eq:5}), which assumes the stream structure to a uniform cylindrical tube, whose density is determined from the maximum shower activity and the cross-sectional area to the ecliptic plane is derived from the shower duration. Consequently, this dust tube model generates an activity profile that looks like a boxcar function with the peak corresponding to the maximum activity. On top of that, an additional multiplication factor $f$ is introduced as well, to account for the more realistic situation where Earth does not traverse the stream's center. For the Geminids, this factor is often chosen as 10. With all of these conditions combined, it is natural that estimates derived from this method would tend to yield larger values for the total Geminid stream mass. 

The idea behind \citet{2017P&SS..143..125R} is akin to our mass estimation method, but the crucial difference lies with the selected dust mass range. As mentioned earlier, we chose the particle size range of 0.1 mm - 10 cm to maintain consistency with the measurements of \citet{2017P&SS..143...83B}, and almost all Geminid mass estimations do not set a lower limit smaller than 0.1 mm. On the other hand, \citet{2017P&SS..143..125R} is an exception and set the lower mass limit as $10^{-12}$ g, roughly equivalent to a particle size of 0.001 mm. Notably, considering that the total mass of particles with radius $r$ is proportional to $r^3 \times r^{-3s-1}$, expanding the mass range to 0.001 mm would increase the total stream mass by at least seven orders of magnitude. In other words, the discrepancy between our mass estimate and that of \citet{2017P&SS..143..125R} largely arises from our use of a narrower mass range. 

This difference stems from how one defines the Geminid stream. In the case of \citet{2017P&SS..143..125R}, the entire dust population released by Phaethon during the formation event is considered to be the Geminid stream. Indeed, particles of $\ge$ 0.001 mm in size can survive being blown away by radiation pressure after being ejected from Phaethon \citep{2021Icar..35413949M}, but the PR lifetime of particles of $\leq$ 0.01 mm is $\lesssim$10\,000 yr \citep{1949AJ.....54R.138W}, which is comparable to the timescale of our simulations. In other words, at the timescale of this study, particles smaller than 0.1 mm ejected directly from Phaethon has deviated significantly from the Geminid shower particles and no longer contributes to the meteor shower. Unlike the observed Geminid meteors, this subset of Phaethon-born dust particles is not constrained by observation and its boundary between the background sporadic complex is vague \citep{2021Icar..35413949M}. On the other hand, our idea of the Geminid stream in this study is the dust complex that is in the vicinity of Earth orbit and is directly connected to the Geminid shower. As we have shown from our results, the stream of mm- and cm-sized particles match this description and therefore, we define the Geminid stream to be the dust stream consisting of these particles and their products of collisonal cascading. Nevertheless, we find both definitions used by \citet{2017P&SS..143..125R} and our study to be valid, but note that the impact of using different definitions is significant for stream mass estimation.

\subsection{Activity profile}
\label{subsec:activity}

In Sect. \ref{subsec:streamloc}, we found several noteworthy models whose peak solar longitude reaches very close to the known maximum activity plateau. We searched for models with (1) peak solar longitude lower than 262.45\degr and (2) more than 10 particles constituting the peak. We set (1) to match the Geminid observation while (2) to ensure statistical reliability. These criteria left us with 2 models:  Clone 50 18000 yr ago and nominal Phaethon 18000 yr ago. Henceforth we will call them Models A and B. Using the Earth-Moon distance criterion, the mass of Models A and B are $1.7 \times 10^{13}$ g and $2.5 \times 10^{12}$ g, respectively, in Fig. \ref{fig:moidmass}. 

\begin{figure}
\resizebox{\hsize}{!}{\includegraphics{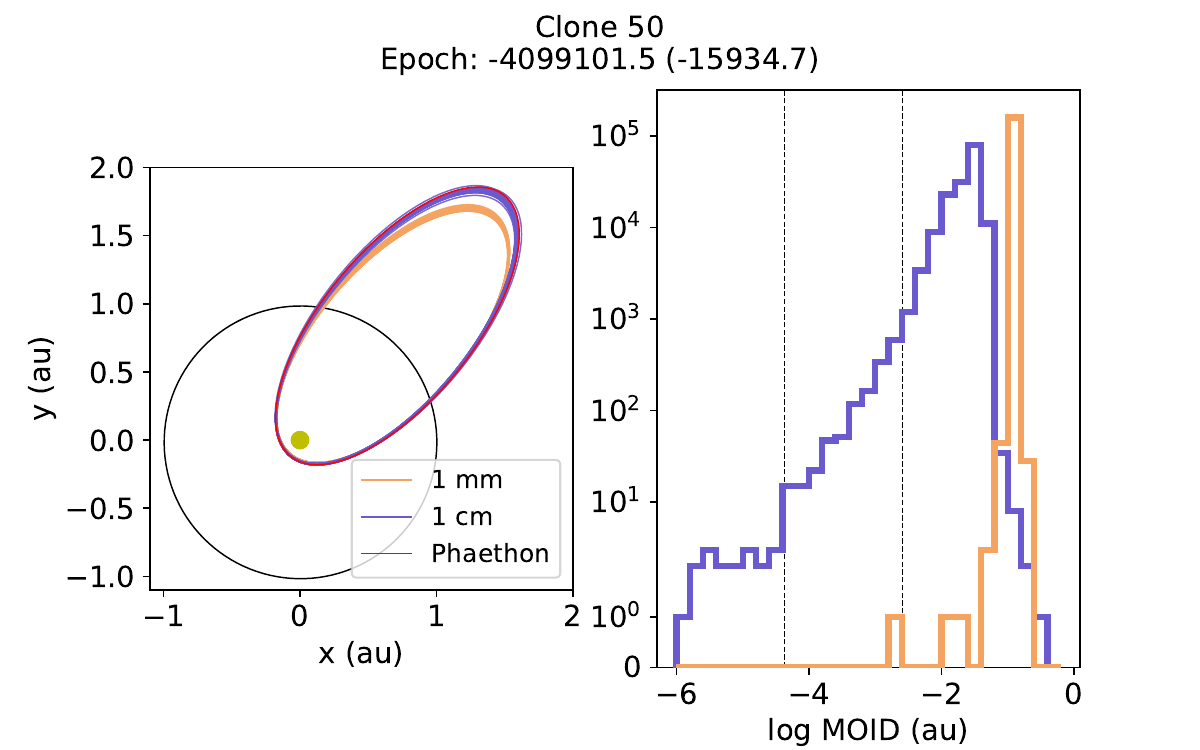}}
\resizebox{\hsize}{!}{\includegraphics{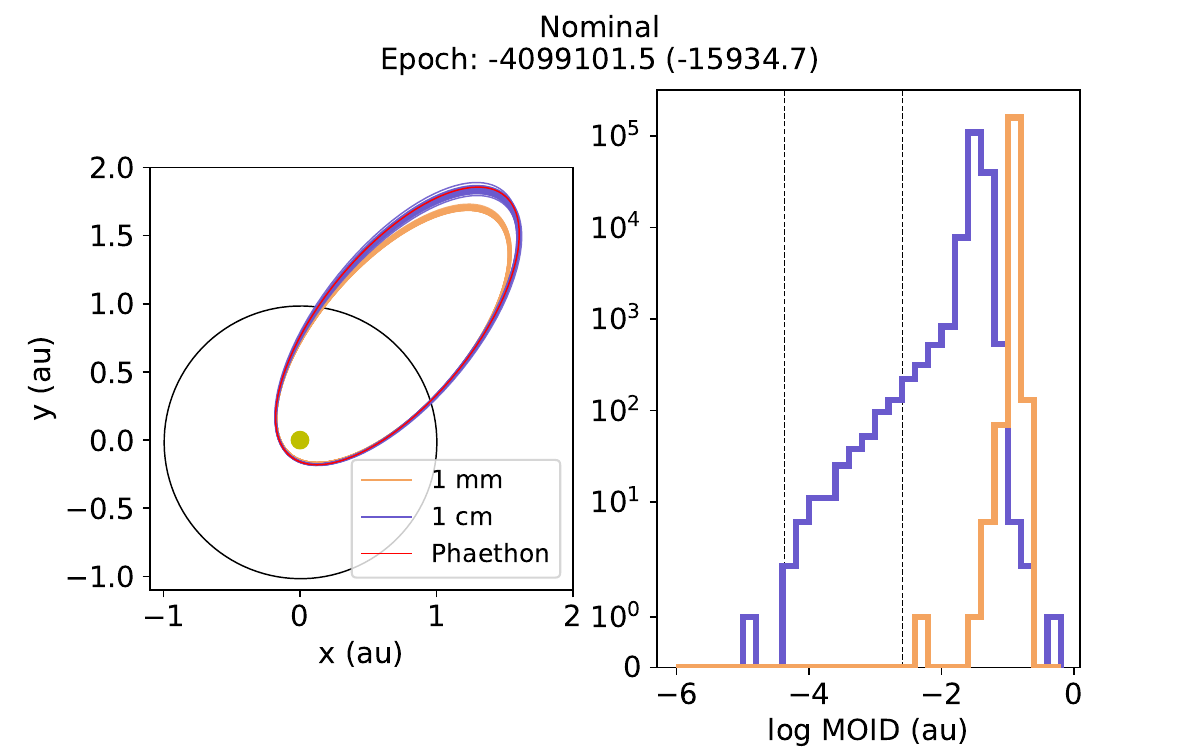}}
\caption{Same as Fig. \ref{fig:nominalhist}, but for Model A (top) and B (bottom).
\label{fig:moidhist_activity}}
\end{figure}

\begin{figure}
 \resizebox{\hsize}{!}{\includegraphics{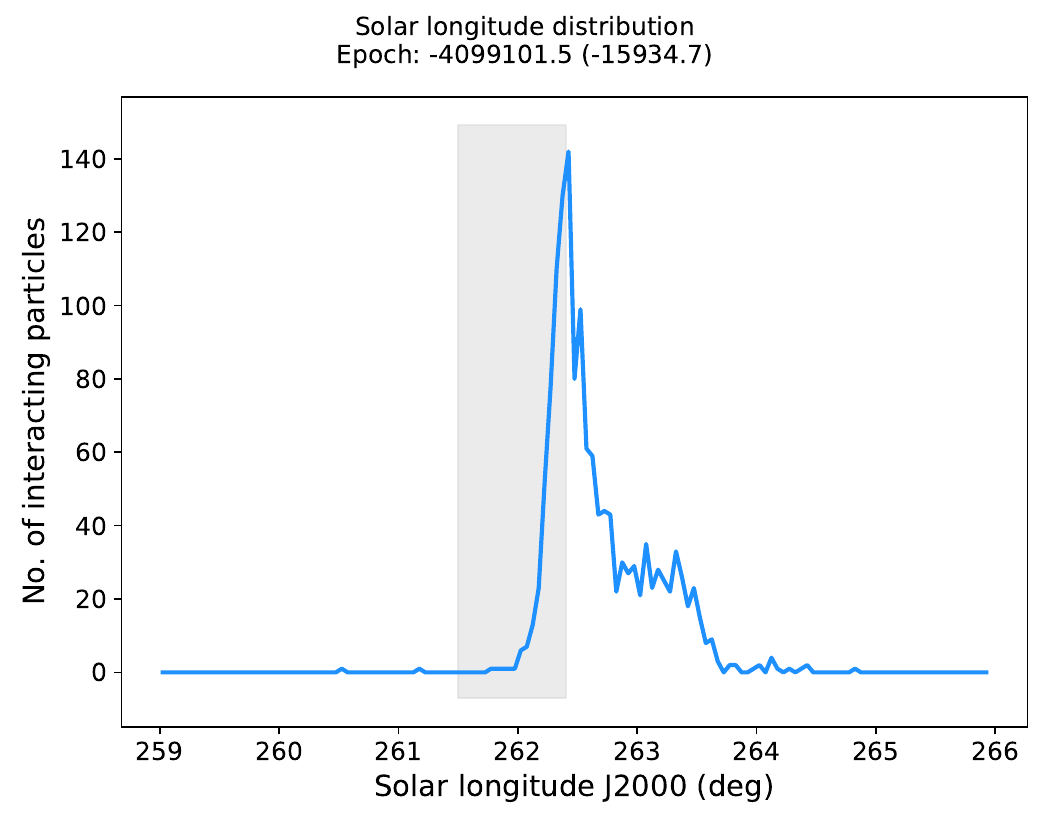}}
 \resizebox{\hsize}{!}{\includegraphics{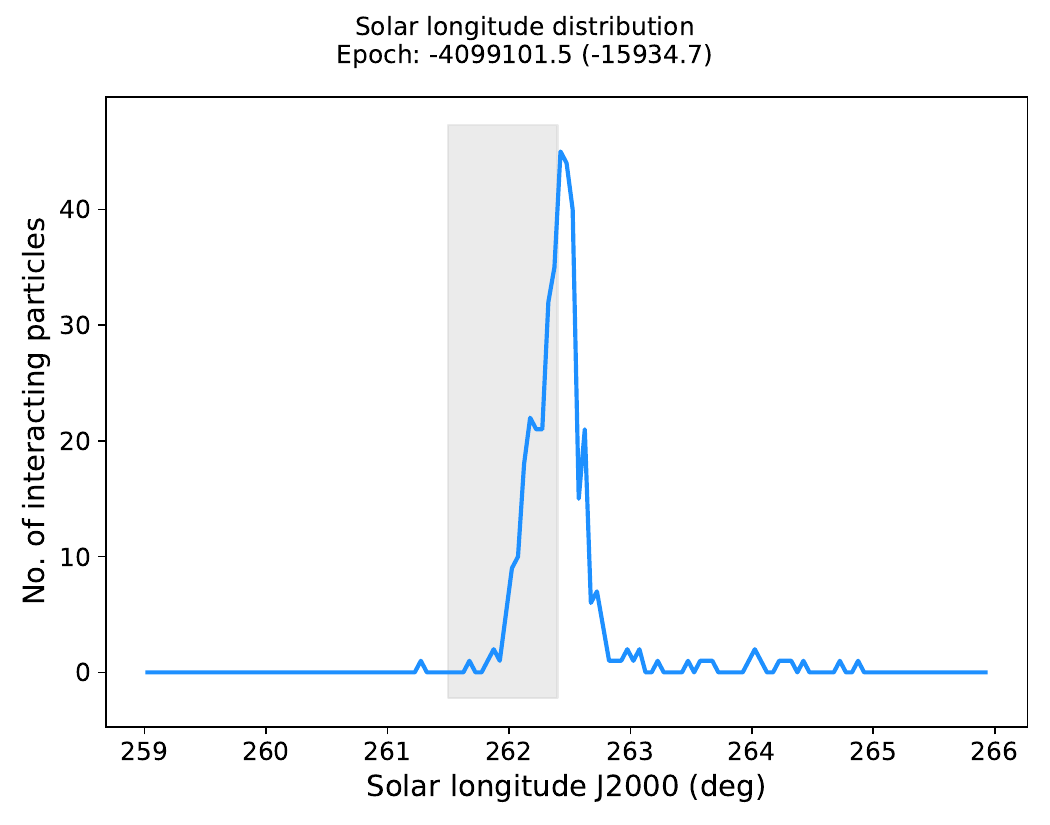}}
\caption{Number density of the points in Fig. \ref{fig:ecliptic} by solar longitude (J2000.0) for Models A (top) and B (bottom). The histogram is binned with an interval of $0.05$\degr. The shadowed area represents the maximum activity plateau by \citet{2004EM&P...95...27R}.
\label{fig:profile}}
\end{figure}

Fig. \ref{fig:profile} presents the number density of 1 cm particles divided into 0.05\degr\ bins of solar longitude. This distribution can be interpreted as the activity profile of the model Geminids observed from Earth. Both Model A and B exhibit a peak at $262.425$\degr. Notably, these locations are near the upper boundary of the maximum activity plateau, $262.4 \pm 0.05$\degr\ \citep{2004EM&P...95...27R}, and imply that larger particles are concentrated at the descending side of the profile, which is in agreement with radio observation data \citep{2009CoSka..39....5Z}. In short, our investigation has led us to create a Geminid stream model that closely approximates the observed peak activity location, despite employing stringent distance conditions to evaluate meteoroid interaction with Earth. However, while our activity profile demonstrates promising aspects, it is not without its limitations, which we will address in Sect. \ref{sec:limits}.

\begin{figure}
\resizebox{\hsize}{!}{\includegraphics{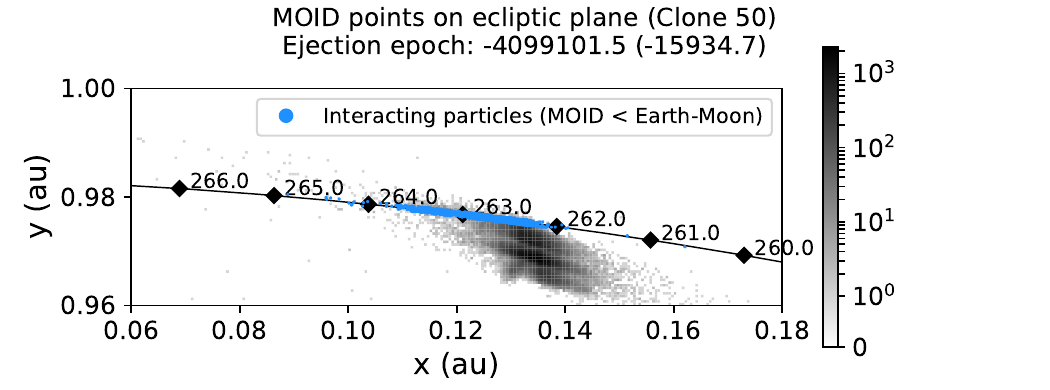}}
\resizebox{\hsize}{!}{\includegraphics{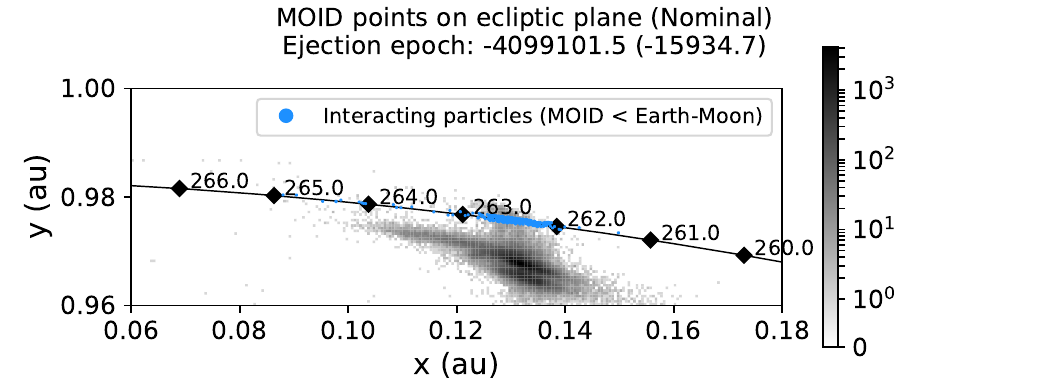}}
\caption{MOID points of the particle orbits with MOID $<$ Earth-Moon distance projected on the ecliptic plane. The reference frame is the Cartesian J2000.0 ecliptic coordinate system and the Earth's orbit is drawn in a solid black line with solar longitudes labeled with diamonds.
\label{fig:ecliptic}}
\end{figure}

Fig. \ref{fig:ecliptic} provides a visual representation of the cross-section of the stream. The core of the stream is $\sim 0.01$ au away from Earth orbit and, compared to the interacting particles, is skewed to lower solar longitude. Thus, as we pointed out earlier in Sect. \ref{subsec:clone}, using $\gtrsim 0.01$ au as the distance condition to determine particle intersection with Earth will overrepresent stream core particles, which can result in misleading statistical analysis.

\subsection{The dynamics of the Geminid meteoroids}
\label{subsec:dynamics}
\begin{figure}
\resizebox{\hsize}{!}{\includegraphics{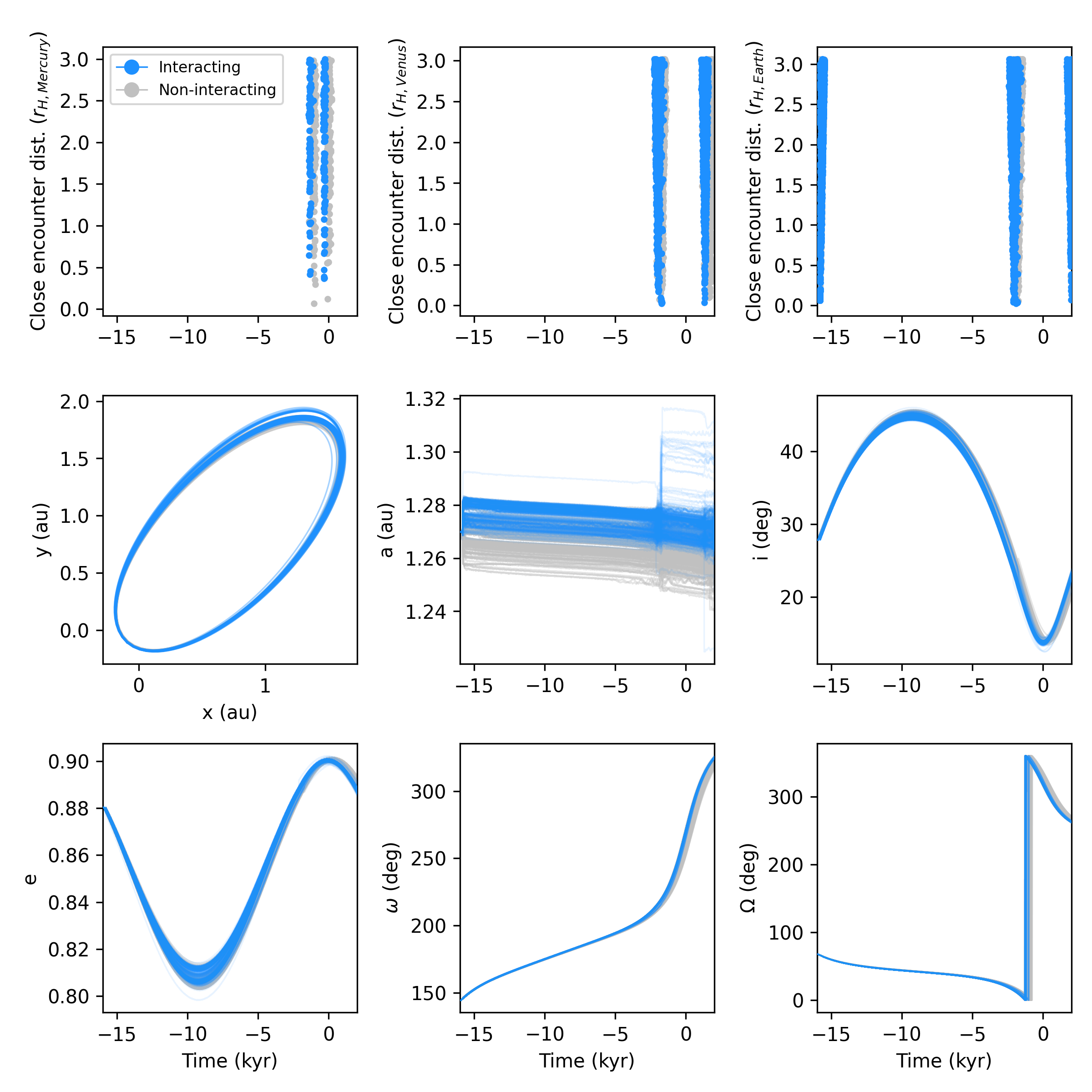}}
\caption{Orbital evolution of Earth-interacting particles (blue) and randomly selected non-interacting particles (gray). The top row displays close encounter distances ($\leq$ 3 Hill radius) of the particles with Mercury, Venus, and Earth throughout the simulation. From middle to bottom row, from left to right, are the present-day orbit of the particles and the time plot of the semimajor axis, inclination, eccentricity, argument of perihelion, and longitude of ascending node.
\label{fig:dustevol}}
\end{figure}

In this section, we aim to identify the differences between particles that interact with Earth and those that do not. Fig. \ref{fig:dustevol} summarizes the orbital histories of interacting and randomly selected non-interacting 1 cm particles from Model B. The distinction between the two particle categories becomes visibly evident in the semimajor axis panel. Here, we observe two notable divergences separating the interacting and non-interacting particles. These divergences occurred  $\sim$18\,000 and $\sim$4000 years ago, coinciding with the close encounter of Venus and Earth. In the case of $\sim$4000 years ago, the particles encountered both major terrestrial planets almost at the same time. Such encounters likely increased the chaotic behavior in the stream's orbit, causing a portion of the stream to migrate to larger semimajor axes. Subsequently, this migration eventually would have led to a section of the stream directly crossing Earth in the present day. Looking back on Fig. \ref{fig:backward}, one can also notice hints of these planetary encounters in the backward integration of Phaethon as well. 

\begin{table*}
\caption{Orbital elements of Earth-interacting particles (JD 2459000.5)}              
\label{table:dusttable}      
\centering                                      
\begin{tabular}{c c c}          
\hline\hline                        
Element  & This work & Hajdukov{\'a} et al. (2017) \\    
\hline                                   
    Perihelion distance, $q$ [au] & $0.143 \pm 0.002$ & $0.147 \pm 0.01$ \\
    Eccentricity, $e$ & $0.887 \pm 0.001$  & $0.884 \pm 0.015$  \\
    Inclination, $i$ [\degr] & $22.959 \pm 0.312$ & $22.94 \pm 1.89$ \\
    Longitude of ascending node, $\Omega$ [\degr]  &  $262.467 \pm 0.438$ & $261.37 \pm 1.62$  \\
    Argument of perihelion, $\omega$ [\degr] & $324.846 \pm 0.319$ & $324.27 \pm 1.40$ \\
\hline                                             
\end{tabular}
\end{table*}

Table \ref{table:dusttable} compares the orbital elements of the Earth-interacting particles with those found by video observations of the Geminid meteoroids \citep{2017P&SS..143...89H}. While all five orbital elements fall within the range of the standard deviation of the observed data, it is important to acknowledge that this is partially attributable to the standard deviation being relatively wide (we will discuss further in Sect. \ref{sec:limits}).

\begin{figure}
\resizebox{\hsize}{!}{\includegraphics{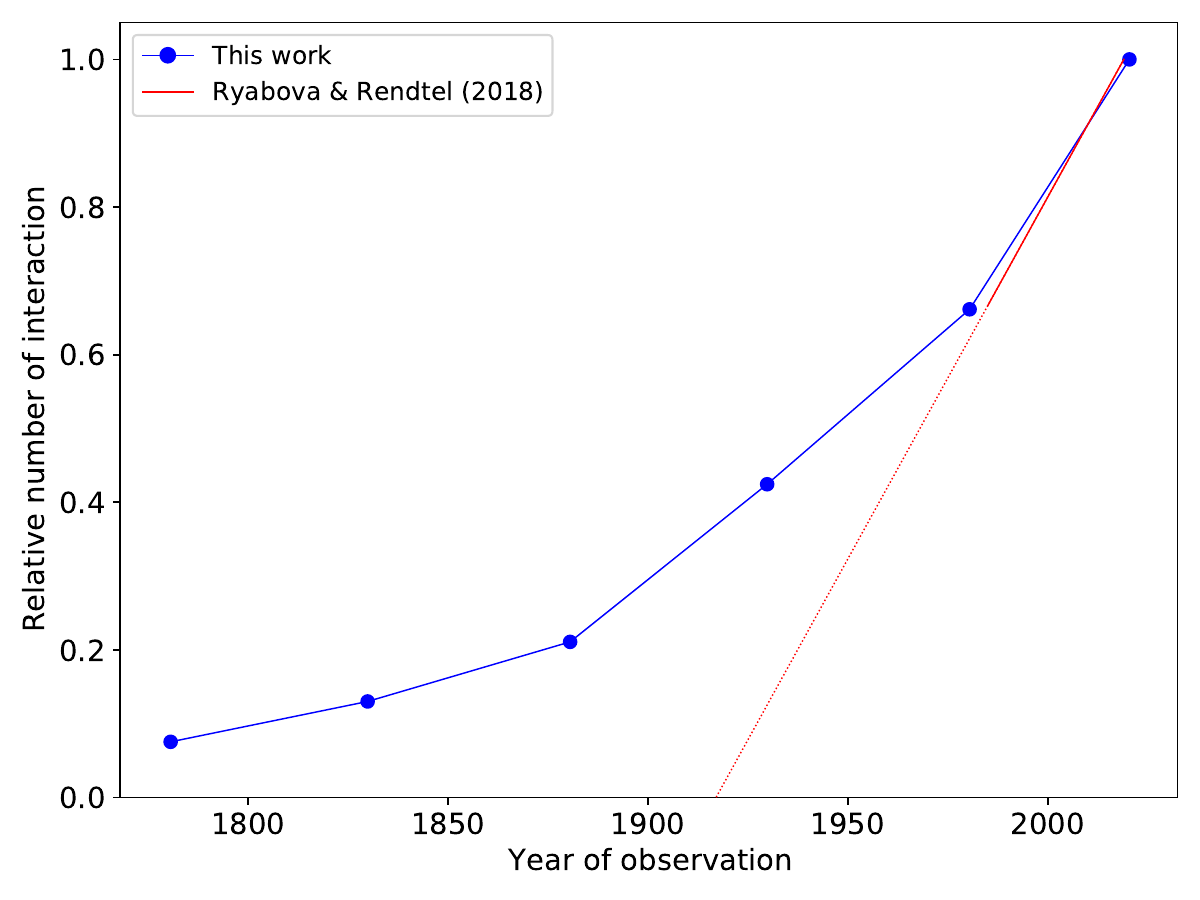}}
\caption{Relative number of interacting particles from Model B in 1780-2019 with a 50-year cadence. The solid red line is the linear fit to the observed activity of 1985-2016 analyzed in \citet{2018MNRAS.475L..77R}, and is extended by the dotted line for visual reference. All $y$-axis values were normalized to the value in 2019.
\label{fig:moidhistory}}
\end{figure}

The first reports of the modern Geminid shower date back to the 19th century, with observational records from the 1880s or even as early as the 1830s in literary accounts \citep{1991BAAS...23..941C,2015MNRAS.453.1186J}. 
Furthermore, \citet{2018MNRAS.475L..77R} showed that the Geminid activity level increased based on numerical simulation and visual observations. We performed a similar analysis in Fig. \ref{fig:moidhistory} and found that our model activity also has been growing for the past two centuries. In particular, the activity increase from the 1980s to the present day is similar to the fit by \citet{2018MNRAS.475L..77R}, albeit slightly shallower.

\subsection{Comparison in the Geminid age with previous research}

Traditionally, the age of the Geminid stream was estimated to be $\sim 1000 - \sim 10000$ years according to the works of \citet{1978MNRAS.183..539J,1985MNRAS.217..523J,1993MNRAS.262..231W,1999SoSyR..33..224R,2002MNRAS.336..559B,2015MNRAS.453.1186J}. More recent studies tended to favor an age of $\sim 2000$ years \citep{2007MNRAS.375.1371R,2016MNRAS.456...78R,2017P&SS..143..125R}. In this context, our work suggests that the Geminid age might be older by several factors or even an entire order of magnitude than these recent estimates. However, we argue that this widely known age estimate should be approached with caution. 

Studies such as \citet{1993MNRAS.262..231W} backward integrated observed Geminid meteoroids or their mean orbital elements, sometimes without applying the Poynting-Robertson effect. \citet{1985MNRAS.217..523J} also utilized backward integration and arrived at the age of 6000 years by matching their result with the observed bimodality. Another approach involves using the secular variation in the meteoroids' orbital elements, assuming that this variation is the combined product of the Poynting-Robertson effect and corpuscular drag \citep{1986AVest..20..142B,1999SoSyR..33..224R}. Yet, as \citet{1999SoSyR..33..224R} pointed out, these methods are highly sensitive to uncertainties in the initial parameters and should be interpreted carefully. 

Another factor that was often underestimated in previous studies was the planetary perturbation of terrestrial planets, especially Venus and Earth. \citet{2007MNRAS.375.1371R} employed the semi-analytical, nested polynomial method to create a model Geminid stream with a presumed age of 2000 years with only Jupiter's perturbation included. Although this method may be valid for 2000 years, as described in Sect. \ref{subsec:dynamics}, the gravitational influences exerted by Venus and Earth become non-negligible when extending the analysis to epochs older than 4000 years. Subsequent numerical simulations, like those of \citet{2016MNRAS.456...78R}, started incorporating these effects as well, but still did not venture beyond the 2000-year mark. 

We only found \citet{2015MNRAS.453.1186J} and \citet{2018MNRAS.479.1017R} to be the studies that probed ejection epochs further than 4000 years ago, while accounting for planetary perturbations and the Poynting-Robertson effect. In the case of \citet{2015MNRAS.453.1186J}, they ejected particles of various $\beta$ values in a sequence of ejection events spanning from $10^3 - 10^5$ yr and compared the orbital elements of their model to the observed meteoroids. Their analysis, based on the longitude of ascending node distribution, indicated that particles with $\beta = 0.009$ launched 1000 yr ago fit well with observed meteoroid orbits. However, one should note that the $\beta$ value is at least one order of magnitude larger than that used in our study. For other epochs, only the orbital elements for the combined stream of all simulated $\beta$ values were provided, rather than detailed results for each specific $\beta$ value, thereby limiting comparison with our findings. On top of that, the selected distance criterion was 0.05 au, which we demonstrated it in Sect. \ref{subsec:activity} to be too lenient. 

Meanwhile, \citet{2018MNRAS.479.1017R} used 100 1-mm-sized particles and ejection epochs of 500 - 5000 years ago with an initial velocity of 30 m/s and the distance criterion of 0.03 au. The author concluded that such ejection did not produce a meteor shower in the present day. However, this may be likely due to the small number of particles in her simulation. As demonstrated by our work (see Figs. \ref{fig:moidnum1mm} and \ref{fig:moidnum1cm}), we find some particles satisfying the MOID criteria at a similar age range using 320\,000 particles, albeit assumed ejection velocities are different from each other. As only a minority of 1-mm particles interacted with Earth in our simulations, we propose that a significant portion of the present-day mm-sized Geminid meteors were generated through collisional cascading of larger particles within this timescale, as discussed in Sect. \ref{subsec:collisional}.

Apart from dynamical simulations, \citet{2002MNRAS.336..559B} theorized that the flickering of Geminid fireballs was caused by their rotation and by assuming that the age of the fireballs is the time it took for solar radiation scattering to spin them up, he derived an age range of 1000-4000 years. This idea was challenged by \citet{2004A&A...428..241B} who instead suggested that the flickering was due to an autofluctuation mechanism of meteoroid ablation.

In short, the Geminid age range that has become widely referred to should be dealt with caution, given that it originates from studies constrained by computational limitations or potentially unreliable methodologies, and since then, this age range has not been properly tested. However, while we acknowledge that even our age estimate of 18\,000 yr is not far off this range, we also encourage future research engaging in numerical simulations of the Geminids to delve into various epochs in detail.

\subsection{Limitations and future work} \label{sec:limits}

Based on stream location, our models have shown a remarkable resemblance to the observed Geminid meteoroids. Nonetheless, there are several limitations of our models that warrant addressing. A prominent Geminid activity feature is the bimodality, or double-peak. Geminid bimodality with a $\sim$ 1\degr gap has been predicted by \citet{1989JIMO...17..240R} and has been consistently observed throughout decades of observations. The observed gap between the two peaks varies from year to year, but generally is around $0.1$\degr \citep{2004EM&P...95...27R}. Although Models A and B both display some minor peaks near the main peak, their magnitudes are insufficient to be interpreted as part of a double peak. However, it is important to note that direct comparison with observation in terms of bimodality can be challenging, as the number, location, spacing, and strength of Geminid activity peaks can vary widely depending on the year of observation \citep{2004EM&P...95...27R,2009CoSka..39....5Z}. 

Another feature that we could not replicate is stream width. The widths of the models' profiles are much narrower than observed activity profiles, which can span up to $\sim 10$\degr. This discrepancy is reflected in the standard deviations of the orbital elements in Table \ref{table:dusttable}, where the observed data clearly shows greater dispersion. Previous simulations, even with significantly higher initial velocities of $\sim 100$ m s\textsuperscript{-1}, also could not reproduce a stream width larger than 3 \degr \citep{2007MNRAS.375.1371R, 2016MNRAS.456...78R}. Therefore, we propose that the initial dispersion during dust ejection is not likely to be the sole contributor to the stream width. Instead, it is crucial to bear in mind that our activity profiles were constructed under the assumption that the entire Geminid stream was created by a single ejection event. However, if the stream was born from multiple ejection events within $<\sim 1000$ years, the combined infall of the ejected particles can look more similar to the multi-peak, widely dispersed activity of the Geminids. For example, one could imagine a scenario where rotational instability is the initial major mass ejection event, subsequently triggering the exposure and volatilization of subsurface sodium sources, which in turn lead to secondary ejection events \citep{2023PSJ.....4...70Z}. Taking into account that Phaethon's current rotational acceleration is faster than that predicted by the YORP effect \citep{2022DPS....5451407M}, one could also conjecture that it could have reached its critical spin period multiple times through a spin up-down-up sequence that occurs more frequently than predicted by the YORP timescale. We also note that the stream width problem was addressed by previous studies such as \citet{2016MNRAS.456...78R}, who propose that the observed width could be caused by Phaethon's orbit being altered during the Geminid formation event. While this idea is plausible for directional ejections, such as comet-like activity, since mass ejection by rotational instability would theoretically be axisymmetric, we consider the orbital change caused by the ejection to be negligible.

Although this work shows the viability of low-velocity dust ejection as the initial dynamical condition of Geminid meteoroids, we leave it to future works to probe whether rotational instability on Phaethon can eject the estimated mass of our models. Furthermore, we anticipate that the upcoming DESTINY\textsuperscript{+} mission will greatly contribute to our understanding of the rotation and activity of Phaethon. As Hayabusa2 and OSIRIS-REx demonstrated from their in-situ observation of Ryugu and Bennu, insights could potentially be derived from assessments of space weathering effect, boulder arrangement, or crater distribution \citep{2019Sci...364..252S, 2020Sci...370.3660D,2020JGRE..12506475J}.

\section{Summary and Conclusions} \label{sec:conclusion}

We presented the first comprehensive investigation of dust ejection on asteroid Phaethon via rotational instability. We conducted numerical simulations of mm- and cm-sized dust ejection $\sim 2 - 100$ kyr ago, assuming a low initial velocity induced by rotational instability. Using strict and realistic MOID criteria, we found particles crossing Earth at present day from multiple ejection epochs. By deriving the location of the peak activity of our models, we found two models that are close to the observed peak solar longitude. We estimated the mass of the model streams to be $\sim 10^{12}$ g, or $\sim 10^{-5}$ Phaethon mass. The models are $\sim 18000$ years old, which is more than 10 kyr older than the previously assumed Geminid age. We analyzed the orbital evolution and the simulation particles and identified close encounters with Venus and Earth to be the key events that caused some of the ejected dust to migrate to Earth-intersecting orbits in the present day. However, our models are not perfect, as we could not replicate the width and bimodality of the Geminid stream.

To summarize, this work demonstrates that Geminid formation by rotational instability is dynamically possible. Further investigation is required to verify that rotational instability can sufficiently supply the estimated mass of our models. We also expect that future observation from the DESTINY\textsuperscript{+} mission will give us valuable insight into the nature of Phaethon's activity and the birth of the Geminids.

\begin{acknowledgements}
This research at SNU was supported by a National Research Foundation of Korea (NRF) grant funded by the Korean government (MEST)
(No. 2023R1A2C1006180). This research utilized the computational facility gmunu and we thank Prof. Hyung Mok Lee, Prof. Woong-Tae Kim, Gain Lee and Da-jung Jang for generously sharing it with us, as well as Dr. Hee-il Kim for kindly providing us with technical support. The authors thank Yoonsoo Bach, Sunho Jin, Bumhoo Lim, and Jooyeon Geem for their helpful comments and discussions as well. We would also like to thank G. O. Ryabova for the insightful review as well.

\end{acknowledgements}

%
%
\bibliographystyle{aa} 
\bibliography{geminid.bib} 

\begin{appendix} 
\section{Clone generation parameters} \label{app:clones}

\begin{table}
\caption{\label{table:covariance}Covariance matrix of the orbital elements of Phaethon at JD 2459000.5}
$\begin{aligned}
\left(\begin{matrix}
  1.797939215310278 \times 10^{-18} &    -2.235936592563117 \times 10^{-18} &   -6.005354667013037 \times 10^{-17} &   2.236255134692548 \times 10^{-16} \\
  -2.235936592563117 \times 10^{-18} &    2.785494863480247 \times 10^{-18} &    8.446162231720073 \times 10^{-17} &   -2.59319466153449 \times 10^{-16} \\
 -6.005354667013037 \times 10^{-17} &    8.446162231720073 \times 10^{-17} &    3.800421272278023 \times 10^{-14} &    2.129774140459485 \times 10^{-14} \\
 2.236255134692548 \times 10^{-16} &    -2.59319466153449 \times 10^{-16} &    2.129774140459485 \times 10^{-14} &    7.985680365362394 \times 10^{-13} \\
 -4.992568539491595 \times 10^{-17} &    3.588252528515852 \times 10^{-17} &    -3.309073480577271 \times 10^{-14} &   -7.731669094340946 \times 10^{-13} \\
  1.08246963991656 \times 10^{-16} &    -1.436062739821552 \times 10^{-16} &    -2.076220487196175 \times 10^{-14} &    -1.753832948981064 \times 10^{-13} \\
  3.603354286149994 \times 10^{-25} &    -4.126543155456637 \times 10^{-25} &   5.709279332176283 \times 10^{-23} &    2.138913369291242 \times 10^{-22}
\end{matrix}\right.\\
\left.\begin{matrix}
  4.992568539491595 \times 10^{-17}  &  1.08246963991656 \times 10^{-16} &   3.603354286149994 \times 10^{-25} \\
  3.588252528515852 \times 10^{-17}   & -1.436062739821552 \times 10^{-16} &    -4.126543155456637 \times 10^{-25} \\
  -3.309073480577271 \times 10^{-14} &    -2.076220487196175 \times 10^{-14} &    5.709279332176283 \times 10^{-23} \\
  -7.731669094340946 \times 10^{-13} &   -1.753832948981064 \times 10^{-13} &   2.138913369291242 \times 10^{-22} \\
  7.866353952006751 \times 10^{-13} &   1.929010520937435 \times 10^{-13} &   -2.30237849060794 \times 10^{-22} \\
  1.929010520937435 \times 10^{-13} &    6.588049822898363 \times 10^{-14} &   -5.008780211368663 \times 10^{-23} \\
  -2.30237849060794 \times 10^{-22} &    -5.008780211368663 \times 10^{-23} &   3.503260996681102 \times 10^{-31}
\end{matrix}\right)
\end{aligned}$.
\end{table}

The covariance matrix of the orbital elements of Phaethon we used is shown in Table \ref{table:covariance}. The rows and columns of the matrix are in the following order: $e$, perihelion distance (au), perihelion epoch (JD), $\Omega, \omega, i, A_2$. As mentioned in Sect. \ref{sec:method}, we generated 100 clones from this matrix. Semimajor axis and mean anomaly, which are needed for Mercury6, were derived from the given orbital elements. Table \ref{table:clone} shows the orbital elements of the clones selected for use in this study. 

   \begin{table}
      \caption{Orbital elements of Phaethon clones used in this work (JD 2459000.5)}
         \label{table:clone}
         \begin{tabular}{l c c c c c}
            \hline
            \noalign{\smallskip}
            Element      &  Clone 72  &  Clone 50 &  Clone 26 & Clone 22 \\
            \noalign{\smallskip}
            \hline
            \noalign{\smallskip}
            $a$ [au] & 1.271367884216 & 1.271367884597 & 1.27136788509 & 1.271367883602\\
            $e$ & 0.8898311239595  &  0.8898311233282 & 0.8898311206132 & 0.8898311204817\\
            $i$ [\degr] & 22.25951208418 & 22.259511909682 & 22.25951174554 & 22.25951161385\\
            $\Omega$ [\degr]  & 265.2176950862 & 265.2176960709 & 265.2176959916 & 265.2176964139\\
            $\omega$ [\degr] & 322.1867110036 & 322.1867099388 & 322.1867096535 & 322.1867093484\\
            Mean anomaly [\degr] & 228.9547681886025 & 228.95476821590543 & 228.95476835484465 & 228.95476816931136\\
            $A_2$ [$10^{-15}$ au/day$^2$] & -5.539668422360 &  -4.957751734560 & -4.740556817135 & -5.85834326728\\
            \hline
         \end{tabular}
   \end{table}
   
\end{appendix}

\end{document}